\documentclass[journal]{IEEEtran}
\IEEEoverridecommandlockouts

\usepackage[utf8]{inputenc}
\usepackage[T1]{fontenc}
\usepackage{textcomp}

\usepackage{graphicx}
\graphicspath{{./figure/}}
\usepackage{amsmath,amssymb,amsfonts}
\usepackage{amsthm}
\usepackage[version=4]{mhchem}
\usepackage{siunitx}
\usepackage{longtable,tabularx}
\usepackage{mathrsfs}
\usepackage{algorithmic}
\usepackage{algorithm}
\usepackage{array}
\usepackage{stfloats}
\usepackage{url}
\usepackage{verbatim}
\usepackage{booktabs}

\usepackage{xcolor}
\usepackage{layout}
\usepackage{hyperref}
\usepackage{float}
\usepackage{fancyhdr}
\usepackage{bm}
\usepackage{subcaption}

\usepackage{tikz}
\usetikzlibrary{arrows.meta,calc}

\definecolor{softgray}{HTML}{F3F4F6}
\definecolor{softpurple}{HTML}{E8DFF0}
\definecolor{plotBlack}{HTML}{101010}
\definecolor{plotGray}{HTML}{8E8E8E}
\colorlet{flowchartAccentColor}{softpurple}
\colorlet{flowchartBlockColor}{softgray}
\colorlet{flowchartBorderColor}{plotBlack}

\newcommand{\setalgonumstyle}{%
    \renewcommand{\theenumi}{\arabic{enumi}}%
    \renewcommand{\labelenumi}{\theenumi.}%
    \renewcommand{\theenumii}{\arabic{enumii}}%
    \renewcommand{\labelenumii}{\theenumi.\theenumii.}%
    \renewcommand{\theenumiii}{\arabic{enumiii}}%
    \renewcommand{\labelenumiii}{\theenumi.\theenumii.\theenumiii.}%
    \renewcommand{\theenumiv}{\arabic{enumiv}}%
    \renewcommand{\labelenumiv}{\theenumi.\theenumii.\theenumiii.\theenumiv.}%
}

\theoremstyle{plain}
\newtheorem{proposition}{Proposition}
\theoremstyle{definition}
\newtheorem{assumption}{Assumption}
\newtheorem{remark}{Remark}

\newcommand{\ALIROD}{\textsf{AL-IROD}}

\newcommand{\N}{\mathbb{N}} 
\newcommand{\Ni}[2]{\N_{#1}^{#2}} 
\newcommand{\UU}{\mathcal{U}} 
\newcommand{\XX}{\mathcal{X}} 
\newcommand{\ZZ}{\mathcal{Z}} 
\newcommand{\YY}{\mathcal{Y}} 

\renewcommand{\labelenumiv}{\theenumi.\theenumii.\theenumiii.\theenumiv.}

\usepackage{enumitem}
\renewcommand{\theenumi}{\arabic{enumi}}
\renewcommand{\labelenumi}{\theenumi.}
\renewcommand{\theenumii}{\arabic{enumii}}
\renewcommand{\labelenumii}{\theenumi.\theenumii.}
\renewcommand{\theenumiii}{\arabic{enumiii}}
\renewcommand{\labelenumiii}{\theenumi.\theenumii.\theenumiii.}
\renewcommand{\theenumiv}{\arabic{enumiv}}
\title{Input Design for Angles-Only Initial Relative Orbit Determination Using Active Learning}

\author{Kui Xie\textsuperscript{1,*}, Giovanni Romagnoli\textsuperscript{2,*}, Giordana Bucchioni\textsuperscript{2}, and Alberto Bemporad\textsuperscript{1}
\thanks{\textsuperscript{1}Kui Xie and Alberto Bemporad are with the IMT School for Advanced Studies Lucca, Piazza San Francesco 19, Lucca, 55100, Italy. Email: \texttt{\scriptsize \{kui.xie,alberto.bemporad\}@imtlucca.it}.}
\thanks{\textsuperscript{2}Giovanni Romagnoli and Giordana Bucchioni are with the Department of Information Engineering at the University of Pisa, Via G. Caruso, 16, Pisa, 56122, Italy. Email: \texttt{\scriptsize \{giovanni.romagnoli@phd.unipi.it, giordana.bucchioni@unipi.it\}}.}
\thanks{\textsuperscript{*}Kui Xie and Giovanni Romagnoli contributed equally to this work.}
\thanks{This work was partially supported by the Italian Ministry of Education and Research (MUR) within the framework of the FoReLab project (Departments of Excellence), by the Italian Space Agency (Agenzia Spaziale Italiana, ASI) under the research project ``DORA - Determinazione Orbitale Relativa Autonoma on-board, con applicazioni terrestri e lunari,'' through agreement no. 2024-35-HH.0 (CUP n. F53C24000360001) between ASI and the University of Pisa, and by the European Union (ERC Advanced Research Grant COMPACT, No. 101141351). Views and opinions expressed are however those of the authors only and do not necessarily reflect those of the European Union or the European Research Council. Neither the European Union nor the granting authority can be held responsible for them.}
}

\begin{document}
\maketitle
\begin{abstract}
Angles-only initial relative orbit determination (IROD) is a challenging estimation problem, as the lack of direct range information results in scale ambiguity and weak practical observability. To address these limitations, this paper introduces an input-design approach to angles-only IROD, in which excitation maneuvers are systematically designed within an active-learning framework. From a dual-control perspective, the resulting control inputs promote exploration of the output space to improve initial-state estimation while accounting for station-keeping requirements and control effort. Angular measurements collected under the resulting maneuver sequence are processed by a batch IROD algorithm extended to handle generic input sequences. A first-order analytical covariance matrix of the initial-state estimate is derived to quantify the uncertainty induced by measurement noise and to support the transition from batch estimation to sequential filtering. 
Numerical simulations and Monte Carlo sensitivity analyses show that the proposed input-design strategy improves initialization accuracy and conditioning compared with baseline excitation methods and supports autonomous rendezvous in the tested scenarios.
\end{abstract}


\section{Introduction}
Reliable state estimation is fundamental to the monitoring and control of nonlinear dynamical systems, and ultimately relies on the ability to extract sufficient information about the system state from the available measurements. Since control inputs affect the system trajectory and, consequently, the resulting measurements, their design can also be exploited to improve the information available for state estimation. In this work, we investigate this input-design perspective in the context of angles-only spacecraft relative navigation.

Autonomous relative navigation is a fundamental technology for safety-critical proximity operations in space, including docking, debris removal, and rendezvous in Earth or cislunar orbits \cite{DBI21}. These operations require the chaser spacecraft to accurately estimate its state relative to a target, typically using measurements provided by on-board sensors.
Existing relative navigation methods rely on a variety of sensing technologies, ranging from active sensors, such as radar and LiDAR, to cooperative GNSS-based solutions and passive vision-based systems, each characterized by different trade-offs in accuracy, autonomy, and hardware complexity~\cite{EH21}.
Among the sensor technologies considered, passive optical cameras offer significant advantages in terms of size, mass, power, and autonomy, making them particularly attractive for small satellites and missions with stringent resource constraints~\cite{DQZ24}. However, a major limitation of optical cameras is that they provide only angular measurements (azimuth and elevation) with respect to the target and therefore lack direct range information. This gives rise to the angles-only initial relative orbit determination (IROD) problem, which is inherently challenging due to the associated observability limitations.

\begin{figure}[!t]
	\centering
	\resizebox{1.0\columnwidth}{!}{
	\begin{tikzpicture}[
		x=1cm,y=-0.85cm,
		block/.style={
			rounded corners=2pt,
			draw=flowchartBorderColor,
			fill=flowchartBlockColor,
			text width=1.5cm,
			minimum height=1cm,
			align=center,
			inner sep=1.5pt,
			font=\normalfont\scriptsize
		},
		controller/.style={
			block,
			draw=flowchartBorderColor,
			fill=flowchartAccentColor
		},
		activebar/.style={
			rounded corners=2pt,
			draw=flowchartBorderColor,
			fill=flowchartAccentColor,
			minimum width=1.28cm,
			minimum height=0.7cm,
			inner sep=0pt
		},
		condition/.style={
			rounded corners=2pt,
			draw=flowchartBorderColor,
			fill=flowchartBlockColor,
			minimum width=2.5cm,
			minimum height=0.75cm,
			align=center,
			inner sep=1pt,
			font=\normalfont\scriptsize
		},
		dashedblock/.style={
			block,
			dashed,
			dash pattern=on 3pt off 3pt
		},
		arr/.style={
			-{Stealth[length=1.8mm,width=1.35mm]},
			line width=0.32pt
		},
		plain/.style={
			line width=0.32pt
		},
		lab/.style={
			font=\scriptsize,
			inner sep=0.8pt
		}
	]

	\coordinate (dataC) at (2.9,0.364);
	\coordinate (dualC) at (1.09,1.818);
	\coordinate (systemC) at (3.636,1.818);
	\coordinate (covC) at (6.182,0.364);
	\coordinate (initC) at (6.182,1.818);
	\coordinate (currentC) at (3.636,2.909);
	\coordinate (condC) at (5.727,4.000);
	\coordinate (rvC) at (2.09,4.000);

	\coordinate (split) at (4.727,0.364);
	\coordinate (pTap) at (7.636,0.364);
	\coordinate (xzero) at (7.273,1.818);
	\coordinate (xkfeed) at (2.182,2.909);
	\coordinate (uTap) at (2.18,1.818);

	\node[block,fill=flowchartAccentColor,minimum height=1cm] (dual) at (dualC) {\ALIROD\\input design  \\ [-0.5pt] \tiny(Sec.~\ref{sec:IROD-OB})};
	\node[block,minimum height=0.7cm] (data) at (dataC) {data\\collection};
	\node[block,minimum height=0.7cm] (system) at (systemC) {system  \\ [-0.5pt] \tiny\eqref{eq:CW_equations_state_space}};
	\node[block,minimum height=1cm] (cov) at (covC) {covariance\\estimation  \\ [-0.5pt] \tiny\eqref{eq:P0_analytical}};
	\node[block,minimum height=1cm] (init) at (initC) {initial state\\reconstruction  \\ [-0.5pt] \tiny\eqref{eq:LOS cond u} \& \eqref{eq:LS problem}};
	\node[block,minimum height=0.7cm] (current) at (currentC) {current state\\reconstruction};

	\node[condition] (cond) at (condC)
		{handover condition  \\ [-0.5pt] \tiny\eqref{eq:trans_cond}};
	\node[dashedblock, minimum height=0.7cm] (rv) at (rvC) {rendezvous  \\ [-0.5pt] \tiny(Sec.~\ref{sec:EKF_MPC})};

	\coordinate (dataInU) at ($(data.south)+(-0.18cm,0)$);
	\coordinate (dataInY) at ($(data.south)+(0.18cm,0)$);

	\draw[arr] (data.west) -- (dual.north |- data.west) -- (dual.north);
	\draw[arr] (system.north) -- (3.636,1.12)
		-- (dataInY |- 3.636,1.12) -- node[lab,right,pos=0.50,xshift=0.05cm] {$\widetilde{\bm{y}}_i$} (dataInY);
	\draw[arr] (dual.east) -- node[lab,below,xshift=-0.1cm, yshift=-0.05cm] {$\bm{u}_i$} (system.west);
	\draw[arr] (uTap) -- (uTap |- 3.136,1.12) -- (dataInU |- 3.136,1.12) -- (dataInU);

	\draw[plain] (data.east) -- (split);
	\draw[arr] (split) -- (cov.west);
	\draw[arr] (split) |- (init.west);

	\draw[arr] (cov.east) -- node[lab,above,xshift=0.25cm] {$\mathbf{P}_{0|i}$} (8.364,0.364);
	\draw[arr] (pTap) -- (pTap |- cond.east) -- (cond.east);
	\draw[arr] (cond.west) -- node[lab,above,font=\normalfont\scriptsize] {Yes} (rv.east);

	\draw[plain] (init.east) -- node[lab,above,xshift=0.80cm] {$\hat{\bm{x}}_{0|i}$} (xzero);
	\draw[arr] (xzero) -- (8.364,1.818);
	\draw[arr] (xzero) -- ++(0,1.091) -- (current.east);

	\draw[plain] (current.west) -- (xkfeed);
	\draw[arr] (xkfeed) -- node[lab,above,xshift=0.3cm] {$\hat{\bm{x}}_{i|i}$} (dual.south |- xkfeed) -- (dual.south);

	\end{tikzpicture}}
	\caption{Flowchart of active-learning-based input design for angles-only IROD. The AL-based IROD framework provides an accurate initial state estimate and its covariance, which quantifies estimation quality and determines the handover to the rendezvous phase.}
	\label{fig:ALforIROD}
\end{figure}
Specifically, angles-only measurements do not guarantee full observability of the relative chaser-target state in the absence of additional information or excitation \cite{WG09}, resulting, under linear unforced dynamics, in an inherent ambiguity in the scale of the relative motion (i.e., the range).
To address this limitation, various strategies have been proposed, including introducing a known camera offset~\cite{KG12}, employing stereo vision systems~\cite{CX11,SCG13}, using target feature recognition, or executing predefined maneuvers~\cite{Cha01}. An alternative approach relies on exploiting the natural relative orbital dynamics (e.g., including perturbations \cite{PGL18} or considering a different set of coordinates \cite{GL17}) to eventually achieve observability over time. The effectiveness and conditioning of these approaches depend on the orbital geometry and observation duration. More generally, these solutions involve different trade-offs: camera offsets and stereo systems increase hardware complexity and have limited effectiveness in certain geometries; feature recognition requires prior knowledge of the target; and predefined maneuvers consume propellant and may not adapt to the specific scenario.

A more principled alternative to heuristic maneuvers is \textit{optimal experiment design} (or \textit{input design}), which seeks to generate excitation signals that maximize the information content of the measured data for parameter or state estimation. In the context of machine learning, this paradigm is referred to as \textit{Active Learning} (AL), where the learner actively queries the environment, here by selecting control inputs to obtain the most informative measurements~\cite{Set12, Bem23c}. Although active learning has been extensively explored for static tasks, its extension to dynamical systems is emerging as a powerful tool for system identification and estimation~\cite{XB24, BGR24}. By formulating maneuver design as an AL task, the chaser can compute excitation sequences that promote exploration of the
angular measurement space while balancing estimation objectives, station-keeping requirements, and control effort.

In this work, we address the angles-only IROD problem by developing an AL-based framework that combines excitation maneuver design,
analytical batch state reconstruction, and uncertainty quantification.
The main contributions are as follows:
($i$) an offline input-design formulation that adapts existing dynamical AL methods~\cite{XB24, BGR24} to angles-only IROD, explicitly accounting for relative orbital dynamics and balancing exploration of the predicted line-of-sight (LOS) measurement space with station-keeping requirements and control effort;
($ii$) an extension of the analytical batch IROD solution in~\cite{GP15} to known impulsive control-input sequences, together with rank and geometric conditions for resolving the scale ambiguity and uniquely reconstructing the initial relative state;
and ($iii$) the derivation of a first-order analytical covariance for the batch estimate under directional LOS measurement noise, providing a quantitative characterization of reconstruction uncertainty and a covariance-based handover criterion.

In our framework, depicted in Fig.~\ref{fig:ALforIROD}, we apply a precomputed excitation sequence to obtain an initial-state estimate and its covariance. Once the handover criterion is satisfied, both are propagated to initialize an extended Kalman filter (EKF) and a model predictive controller (MPC) for rendezvous. Monte Carlo sensitivity analyses and closed-loop simulations evaluate estimation accuracy and subsequent navigation performance.

The remainder of the paper is organized as follows.
Section~\ref{sec:IROD} formulates the angles-only IROD problem, presents its batch least-squares solution, and establishes conditions for unique reconstruction.
Section~\ref{sec:covariance_analysis} derives the analytical covariance and introduces the covariance-based handover criterion.
Section~\ref{sec:IROD-OB} describes the AL-based input-design formulation.
Section~\ref{sec:EKF_MPC} presents the EKF--MPC rendezvous architecture.
Section~\ref{sec:num-exp} reports the numerical results, including Monte Carlo sensitivity analyses and closed-loop rendezvous simulations, and Section~\ref{sec:cons} concludes the paper.

\begin{figure}[t!] 
	\centering
	\resizebox{\columnwidth}{!}{%
		\begingroup
		\begin{tikzpicture}[
		x=1.116cm,y=0.88cm,
		axis/.style={draw=black,line width=0.9pt,-{Stealth[length=2.6mm,width=1.8mm]}},
		bodyaxis/.style={draw=black,line width=0.9pt,-{Stealth[length=2.4mm,width=1.7mm]}},
		construction/.style={draw=black,dashed,line width=0.65pt},
		vector/.style={draw=black,line width=1.0pt,-{Stealth[length=2.5mm,width=1.8mm]}},
		angle/.style={draw=black,line width=0.9pt,-{Stealth[length=2.3mm,width=1.6mm]}},
		mathlabel/.style={font=\large},
		textlabel/.style={font=\large},
		]
		\coordinate (O) at (3.05,2.70);
		\coordinate (C) at (9.25,7.55);
		\coordinate (P) at (9.25,0.55);

		\draw[axis] (0.35,2.70) -- (14.55,2.70);
		\draw[axis] (3.05,0.25) -- (3.05,10.20);
		\draw[axis] (0.50,1.45) -- (13.95,7.35);

		\node[mathlabel,anchor=south east] at (2.78,9.72)
		{$z_{\scriptscriptstyle\mathrm{LVLH}}$};
		\node[mathlabel,anchor=south west] at (13.58,7.48)
		{$y_{\scriptscriptstyle\mathrm{LVLH}}$};
		\node[mathlabel,anchor=south east] at (14.43,2.88)
		{$x_{\scriptscriptstyle\mathrm{LVLH}}$};
		\node[textlabel,anchor=north east] at (14.45,2.52) {V-bar direction};

		\draw[construction] (C) -- (P);
		\draw[construction] (O) -- (P);

		\coordinate (L) at ($(O)!0.48!(C)$);
		\draw[vector] ($(O)+(0.30,0.23)$) -- (L);
		\draw[construction] (L) -- ($(C)!0.14!(O)$);
		\node[mathlabel,anchor=south east] at ($(O)!0.31!(C)+(-0.18,0.18)$)
		{$\bm{LOS}$};
		\node[mathlabel,anchor=south east] at ($(O)!0.64!(C)+(-0.10,0.20)$)
		{$\bm r$};

		\draw[angle] ($(O)+(-17.8:1.72)$)
		arc[start angle=-17.8,end angle=38.1,radius=1.72];
		\node[mathlabel] at (5.02,3.05) {$\beta$};
		\draw[angle] ($(O)+(20.8:3.05)$)
		arc[start angle=20.8,end angle=-17.8,radius=3.05];
		\node[mathlabel] at (6.4,3.05) {$\alpha$};

		\draw[bodyaxis] (C) -- ($(C)!0.25!(O)$);
		\draw[bodyaxis] (C) -- (8.00,9.02);
		\draw[bodyaxis] (C) -- (8.95,9.18);
		\node[mathlabel,anchor=south east] at ($(C)!0.23!(O)+(-0.02,0.10)$) {$x_B$};
		\node[mathlabel,anchor=south east] at (8.00,9.05) {$y_B$};
		\node[mathlabel,anchor=south west] at (9.08,9.05) {$z_B$};

		\path[draw=black,fill=plotGray,line width=0.8pt]
		(2.65,2.30) rectangle (3.45,3.10);
		\path[draw=black,fill=gray!62,line width=0.8pt]
		(3.45,2.30) -- (3.73,2.58) -- (3.73,3.38) -- (3.45,3.10) -- cycle;
		\path[draw=black,fill=gray!52,line width=0.8pt]
		(2.65,3.10) -- (2.93,3.38) -- (3.73,3.38) -- (3.45,3.10) -- cycle;
		\node[textlabel,anchor=south east] at (2.50,2.90) {Target};

		\begin{scope}[shift={(C)},rotate=38.1]
		\path[draw=black,fill=plotGray,line width=0.9pt]
			(-0.385,-0.34) -- (0.385,-0.34)
			arc[start angle=-90,end angle=90,x radius=0.18,y radius=0.34]
			-- (-0.385,0.34)
			arc[start angle=90,end angle=270,x radius=0.18,y radius=0.34]
			-- cycle;
		\path[draw=black,fill=gray!62,line width=0.9pt]
			(-0.385,0) ellipse[x radius=0.18,y radius=0.34];
		\end{scope}
		\node[textlabel,anchor=west] at (10.30,8.18) {Chaser};

		\path[use as bounding box] (0.20,0.15) rectangle (14.70,10.35);
		\end{tikzpicture}
		\endgroup
	}
	\caption{Main reference frames and LOS angles. The chaser is equipped with a camera that measures the azimuth ($\alpha$) and elevation ($\beta$) angles to the target, which are used to derive the LOS vector in the LVLH frame.}
	\label{fig:frames_and_angles}
\end{figure}

\section{Problem Formulation and Observability of Angles-Only IROD}
\label{sec:IROD}

We consider the reconstruction of an unknown initial state
from sampled measurements of a controlled system with linear
dynamics and a nonlinear measurement model:
\begin{equation}\label{eq:general_system}
	\begin{aligned}
		\dot{\bm{x}}(t) &= \mathbf{A}\bm{x}(t)
		+ \mathbf{B}\bm{u}(t),\\
		\bm{y}_i &= h(\bm{x}(t_i)),
	\end{aligned}
\end{equation}
where $\mathbf{A}$ and $\mathbf{B}$ are known constant matrices, and
$\bm{u}(t)$ is the applied input. The estimation objective is to recover
$\bm{x}(t_0)$ from the available measurements and known input history. 

In this work, this problem takes the form of angles-only
initial relative orbit determination. A chaser spacecraft,
equipped with a passive optical camera, observes the angular
position of a moving non-cooperative target. The unknown initial state
consists of the relative position $\bm{r}(t_0)\in\mathbb{R}^3$ and relative velocity
$\bm{v}(t_0)\in\mathbb{R}^3$, collected in
$\bm{x}(t_0)=[\bm{r}(t_0)^\top\ \bm{v}(t_0)^\top]^\top\in\mathbb{R}^6$.
Below, we derive a model of the relative dynamics and
measurement geometry used for initial state reconstruction.

\begin{assumption}[Reference orbit and relative motion]
	\label{ass:cw_model}
	The target follows a circular reference orbit of radius $a$,
	and the relative separation satisfies
	$\|\bm{r}(t)\|_2/a\ll1$ over the observation interval.
	The relative dynamics are approximated by the
	Clohessy--Wiltshire (CW) model.
\end{assumption}

The CW model is expressed in the target-centered
local-vertical/local-horizontal (LVLH) frame. Its $z$-axis
is aligned with the target's inertial position vector, its
$y$-axis with the orbital angular momentum, and its $x$-axis
completes the orthogonal triad shown in
Fig.~\ref{fig:frames_and_angles}.
With
$\bm{x}=[\bm{r}^\top\ \bm{v}^\top]^\top
=[x\ y\ z\ \dot{x}\ \dot{y}\ \dot{z}]^\top$,
the matrices in~\eqref{eq:general_system} are
\begin{equation}
	\label{eq:CW_equations_state_space}
	\mathbf{A} =
	\begin{bmatrix}
		\mathbf{0}_3 & \mathbf{I}_3\\
		\begin{smallmatrix}
			0 & 0 & 0\\
			0 & -n^2 & 0\\
			0 & 0 & 3n^2
		\end{smallmatrix}
		&
		\begin{smallmatrix}
			0 & 0 & -2n\\
			0 & 0 & 0\\
			2n & 0 & 0
		\end{smallmatrix}
	\end{bmatrix},\\
	\quad
	\mathbf{B} =
	\begin{bmatrix}
		\mathbf{0}_3\\
		\mathbf{I}_3
	\end{bmatrix}.
\end{equation}
Here $n=\sqrt{\mu/a^3}$ is the target's mean motion,
$\mu$ is the gravitational parameter, and
$\bm{u}(t)=[u_x(t)\ u_y(t)\ u_z(t)]^\top$ is the commanded
acceleration expressed in the LVLH frame.

The state evolution is
\begin{equation}
	\label{eq:conv_integral}
\bm{x}(t)=\bm{\Phi}(t-t_0)\bm{x}(t_0)
+\int_{t_0}^{t}\bm{\Phi}(t-\xi)\mathbf{B}\bm{u}(\xi)
\,d\xi,
\end{equation}
where $\bm{\Phi}(t)=e^{\mathbf{A}t}$ is the state transition
matrix. Partitioning it according to position and velocity
gives the blocks $\bm{\Phi}_{rr}$, $\bm{\Phi}_{rv}$,
$\bm{\Phi}_{vr}$, and $\bm{\Phi}_{vv}$, whose explicit
expressions are provided in~\cite{GP15}.

\begin{assumption}[Measurement geometry and sampling]
	\label{ass:geometry_sampling}
	The camera-to-LVLH rotation is known at each observation instant.
	Measurements are collected at equally spaced times
	$t_i=t_0+iT,  i=0,\ldots,N-1,$
	with $T>0$ and $\|\bm r(t_i)\|_2>0$. Each measurement is acquired
	immediately before the corresponding impulse; no impulse is applied
	at $t_0$. The sampling interval satisfies
	$\det\!\bigl(\bm{\Phi}_{rv}(T)\bigr)\ne0.$
\end{assumption}

For a general input $\bm{u}(t)$, evaluating \eqref{eq:conv_integral} requires computing the convolution integral. To obtain a finite-sum expression for the state evolution, we consider the input as a variable-amplitude Dirac comb with a period $T$ (i.e., $\bm{u}(t) = \sum_{j=1}^{N_u-1} \bm{u}_j \delta(t - (t_0 + jT))$). The convolution integral simplifies into a discrete series, where $\bm{u}_j$ is the amplitude of the $j$-th impulse. Here $N_u$ is the length of $\{\bm{u}_j\}_{j=0}^{N_u-1}$, including $\bm{u}_0=\bm{0}$; inputs beyond this sequence are zero. Defining
$\mathcal{J}(t) =
\{j\in\{1,\ldots,N_u-1\}:t_0+jT<t\},$
the state evolution is
\begin{equation}
	\label{eq:CW_r_v}
	\begin{aligned}
		\bm{r}(t)
		&=
		\bm{\Phi}_{rr}(t-t_0)\bm{r}(t_0)
		+\bm{\Phi}_{rv}(t-t_0)\bm{v}(t_0)\\
		&\quad+
		\sum_{j\in\mathcal{J}(t)}
		\bm{\Phi}_{rv}(t-t_0-jT)\bm{u}_j,\\
		\bm{v}(t)
		&=
		\bm{\Phi}_{vr}(t-t_0)\bm{r}(t_0)
		+\bm{\Phi}_{vv}(t-t_0)\bm{v}(t_0)\\
		&\quad+
		\sum_{j\in\mathcal{J}(t)}
		\bm{\Phi}_{vv}(t-t_0-jT)\bm{u}_j.
	\end{aligned}
\end{equation}
The measurement at each sampling instant is acquired before the
corresponding impulse is applied. Therefore, at $t_i=t_0+iT$,
$\mathcal{J}(t_i)=\{j:1\le j<N_u,\ j<i\}$.
At time $t_i$, the available batch contains $N=i+1$ observations.

Under Assumption~\ref{ass:geometry_sampling}, the ideal camera measurement
can be expressed as a unit LOS vector in the LVLH frame through the
known camera-to-LVLH rotation. Let $\alpha_i$ and $\beta_i$ denote the azimuth and elevation
angles of the LOS in the LVLH frame at $t_i$, respectively. Then
\begin{equation}
	\label{eq:LOS_azel}
	\bm{y}_i =
	\begin{bmatrix}
		\cos \beta_i \cos \alpha_i \\
		\cos \beta_i \sin \alpha_i \\
		\sin \beta_i
	\end{bmatrix}.
\end{equation}
Equivalently, the LOS defines the measurement function
\begin{equation}
	\label{eq:LOS}
	\bm{y}_i
	=
	h\!\left(\bm x(t_i)\right)
	=
	\frac{\bm r(t_i)}
	{\left\lVert\bm r(t_i)\right\rVert_2}.
\end{equation}
The measured LOS is denoted by $\widetilde{\bm{y}}_i$, with a measurement error $\bm{z}_i=\widetilde{\bm{y}}_i-\bm{y}_i$.
Equation~\eqref{eq:LOS_azel} also applies to measured angles and the
measured LOS when all three quantities carry a tilde.
For clarity, Fig.~\ref{fig:frames_and_angles} illustrates the LOS
angles and the main reference frames adopted in this work.

As observed in~\cite{GP15}, the time history of the above LOS measurements is unchanged when the initial conditions are multiplied by an arbitrary scalar $k$. Using~\eqref{eq:CW_r_v} and considering the first three measurements, we have the following equations:
\vspace{-12pt}
\begin{subequations}\label{eq:LOS cond u}
	\begin{align}
		k_0 \bm{y}_0 &= \bm{r}(t_0), \label{eq:LOS cond 1 u} \\
		k_1 \bm{y}_1 &= \bm{\Phi}_{rr}(T)\bm{r}(t_0) + \bm{\Phi}_{rv}(T)\bm{v}(t_0), \label{eq:LOS cond 2 u} \\
		k_2 \bm{y}_2 &= \bm{\Phi}_{rr}(2T)\bm{r}(t_0) + \bm{\Phi}_{rv}(2T)\bm{v}(t_0) + \bm{\Phi}_{rv}(T) \bm{u}_1. \label{eq:LOS cond 3 u}
	\end{align}
\end{subequations}
Assumption~\ref{ass:geometry_sampling} is adopted here only to streamline the derivation; the batch IROD construction can be extended to nonuniform sampling and noncoincident measurement and actuation schedules.

By solving for \( \bm{r}(t_0) \) from \eqref{eq:LOS cond 1 u}, substituting it into \eqref{eq:LOS cond 2 u} to obtain \( \bm{v}(t_0) \), and then substituting both \( \bm{r}(t_0) \) and \( \bm{v}(t_0) \) into \eqref{eq:LOS cond 3 u}, we reduce the system to a set of three equations in the unknowns \( k_0 \), \( k_1 \), and \( k_2 \), with an explicit dependence on the impulsive input:
\begin{equation} \label{eq:syst_u}
	\begin{aligned}
		k_2 \bm{y}_2 = \, & k_0\bm{\Phi}_{rr}(2T)\bm{y}_0 
		+ \bm{\Phi}_{rv}(2T) \bm{\Phi}_{rv}^{-1}(T) \\
		& \cdot \left[k_1\bm{y}_1 - k_0 \bm{\Phi}_{rr}(T)\bm{y}_0\right]
		+ \bm{\Phi}_{rv}(T)\bm{u}_1.
	\end{aligned}
\end{equation}
Solving~\eqref{eq:syst_u} for the scale factors $k_0, k_1, k_2$ determines $\bm{r}(t_0)$ and $\bm{v}(t_0)$ uniquely from \eqref{eq:LOS cond 1 u} and \eqref{eq:LOS cond 2 u}. For $N\ge3$ observations, we formulate a least-squares problem where the $i$-th observation yields:
\begin{equation} \label{eq:LOS cond i u final}
	\begin{aligned}
		k_i \bm{y}_i =\, & k_0\bm{\Phi}_{rr}(iT)\bm{y}_0 + \bm{\Phi}_{rv}(iT)\bm{\Phi}_{rv}(T)^{-1} \\
		& \cdot [k_1\bm{y}_1 -k_0 \bm{\Phi}_{rr}(T)\bm{y}_0] + \sum_{j=1}^{i}\bm{\Phi}_{rv}((i-j)T) \bm{u}_j.
	\end{aligned}
\end{equation}
Finally, it is possible to arrange the system in the form 
$\mathbf{A}_N \bm{k}_N = \mathbf{B}_N$, with $\bm{k}_N = \left[k_0 \ \ldots \ k_{N-1}\right]^\top$, where $N$ represents the number of observations ($N=N_u+1$ when the complete input horizon is observed) and the matrices $\mathbf{A}_N$ and $\mathbf{B}_N$ are defined by~\eqref{eq:A LS} and~\eqref{eq:B LS}, respectively.
For compactness, define
\begin{subequations}
\begin{align}
\mathbf{c}_m &\triangleq \bm{\Phi}_{rr}(mT)\bm{y}_0
- \bm{\Phi}_{rv}(mT)\bm{\Phi}_{rv}(T)^{-1}\bm{\Phi}_{rr}(T)\bm{y}_0, \nonumber \\
\mathbf{d}_m &\triangleq \bm{\Phi}_{rv}(mT)\bm{\Phi}_{rv}(T)^{-1}\bm{y}_1,
\quad m = 2,\ldots,N-1. \nonumber \\[6pt]
\mathbf{A}_N &=
\begin{bmatrix}
\mathbf{c}_2 & \mathbf{d}_2 & -\bm{y}_2 & \cdots & 0 \\
\vdots & \vdots & \vdots & \ddots & \vdots \\
\mathbf{c}_{N-1} & \mathbf{d}_{N-1} & 0 & \cdots & -\bm{y}_{N-1}
\end{bmatrix},
\label{eq:A LS}\\
\mathbf{B}_N &=
\begin{bmatrix}
-\bm{\Phi}_{rv}(T)\bm{u}_1 \\
\vdots \\
-\sum_{j=1}^{N-1}\bm{\Phi}_{rv}((N-1-j)T)\bm{u}_j
\end{bmatrix}.
\label{eq:B LS}
\end{align}
\end{subequations}

With noisy observations, let $\widetilde{\mathbf{A}}_N$ be obtained from~\eqref{eq:A LS} by replacing each $\bm{y}_i$ with $\widetilde{\bm{y}}_i$. For $N\ge3$ and $\operatorname{rank}(\widetilde{\mathbf{A}}_N)=N$, the least-squares estimate of the scale vector is 
\begin{equation}
	\label{eq:LS problem}
	\hat{\bm{k}}_N = \left( \widetilde{\mathbf{A}}_N^\top \widetilde{\mathbf{A}}_N \right)^{-1} \widetilde{\mathbf{A}}_N^\top \mathbf{B}_N.
\end{equation}

Finally, given \eqref{eq:LOS cond u} and $\hat{\bm{k}}_N$ from~\eqref{eq:LS problem}, the initial state $\hat{\bm{x}}_{0|N-1} = [\hat{\bm{r}}(t_0)^\top \ \hat{\bm{v}}(t_0)^\top]^\top$ is reconstructed as:
\begin{equation}
	\label{eq:state_reconstruction}
	\hat{\bm{x}}_{0|N-1} = \widetilde{\mathbf{C}} \mathbf{J} \hat{\bm{k}}_N,
\end{equation}
where $\mathbf{J} = [\,\mathbf{I}_2 \ \ \mathbf{0}_{2\times (N-2)}\,]$ extracts the first two coefficients, $\hat k_0$ and $\hat k_1$, required for the state computation. The matrix $\widetilde{\mathbf{C}}$ is obtained by replacing the ideal LOS vectors in $\mathbf{C}$ with their measured counterparts, where
\begin{equation}
	\mathbf{C} =
	\begin{bmatrix}
		\bm{y}_0 & \mathbf{0}_{3\times 1} \\
		-\bm{\Phi}_{rv}^{-1}(T)\bm{\Phi}_{rr}(T)\bm{y}_0 &
		\bm{\Phi}_{rv}^{-1}(T)\bm{y}_1
	\end{bmatrix}.
\end{equation}

\subsection{Observability Analysis with Impulsive Input}
\label{subsec:observability_analysis}
In the absence of control inputs, angles-only measurements are invariant to positive scaling of the initial state, rendering the unforced system unobservable due to intrinsic scale ambiguity~\cite{WG09}. To restore observability, the known impulsive control input $\bm{u}_1$ introduced in~\eqref{eq:LOS cond 3 u} is exploited to resolve this ambiguity.

Define the auxiliary matrices $\mathbf{M}_0 \triangleq \bm{\Phi}_{rr}(2T) - \bm{\Phi}_{rv}(2T)\bm{\Phi}_{rv}(T)^{-1}\bm{\Phi}_{rr}(T)$ and
$\mathbf{M}_1 \triangleq \bm{\Phi}_{rv}(2T)\bm{\Phi}_{rv}(T)^{-1}$.
Then the consistency relation in~\eqref{eq:syst_u} can be rewritten as
\begin{equation}
	\label{eq:observability_compact}
	k_2 \bm{y}_2 - k_1 \mathbf{M}_1 \bm{y}_1 - k_0 \mathbf{M}_0 \bm{y}_0 = \bm{\Phi}_{rv}(T)\bm{u}_1.
\end{equation}

\begin{proposition}
\label{prop:observability}
Consider the angles-only IROD problem with three ideal LOS vectors and one known impulsive input $\bm{u}_1$. Let
\begin{equation}
\mathcal{O} = \left[ \mathbf{M}_0 \bm{y}_0 \;\; \mathbf{M}_1 \bm{y}_1 \; -\bm{y}_2 \right].
\end{equation}
The scale vector $\bm{k} = [k_0\; k_1\; k_2]^\top$ is uniquely determined by~\eqref{eq:observability_compact} if and only if $\mathcal{O} $ is nonsingular. If $\mathbf{M}_0\bm{y}_0$ and $\mathbf{M}_1\bm{y}_1$ are linearly independent, this is equivalent to
\begin{equation}
	\label{eq:observability_condition}
	\bm{u}_1 \notin \operatorname{span} \left\{ \bm{\Phi}_{rv}(T)^{-1} \mathbf{M}_0 \bm{y}_0,\; \bm{\Phi}_{rv}(T)^{-1} \mathbf{M}_1 \bm{y}_1 \right\}.
\end{equation}
\end{proposition}

\begin{proof}
Equation~\eqref{eq:observability_compact} is linear in the unknown scale coefficients. Hence, $\bm{k}$ is uniquely recoverable if and only if the coefficient matrix $\mathcal{O} $ is invertible. Rewriting~\eqref{eq:observability_compact} as $\mathcal{O} \bm{k} = -\bm{\Phi}_{rv}(T)\bm{u}_1$ shows, for linearly independent first two columns and $k_2>0$, that singularity of $\mathcal{O} $ is equivalent to the right-hand side belonging to the span generated by the first two transformed LOS terms, which yields~\eqref{eq:observability_condition}.
\end{proof}

\begin{remark}
Proposition~\ref{prop:observability} gives a geometric interpretation of the maneuver design problem: the impulsive input must excite a direction that is not already generated by the unforced LOS evolution. In this sense, the maneuver acts as the mechanism that fixes the scale of the relative orbit.
\end{remark}

While Proposition~\ref{prop:observability} is derived for a single impulse, the same logic extends to longer measurement and input sequences. In practice, observability is recovered once the cumulative maneuver effect injects information outside the subspace induced by the unforced dynamics, so that the least-squares problem in~\eqref{eq:LS problem} admits a unique solution.
This observation motivates the active-learning formulation of Sec.~\ref{sec:IROD-OB}: the role of input design is not merely to perturb the trajectory, but to select maneuvers that improve the conditioning of the IROD reconstruction.

\section{Analytical Covariance of the Batch IROD Estimate}
\label{sec:covariance_analysis}
The IROD algorithm described previously provides a deterministic estimate based on a set of noisy observations. To assess the reliability of this solution, it is necessary to quantify how measurement uncertainties propagate through the least-squares inversion to the initial state. Consequently, the objective of this section is to derive the analytical expression for the estimation error covariance under the assumption that sensor noise constitutes the primary source of uncertainty.
\subsection{Angular Measurement Model}
\label{subsec:error model}
\begin{assumption}
	\label{ass:los_noise}
	Each LOS measurement is a unit vector affected by a small angular perturbation that is purely directional and isotropic in the image plane. Specifically, the noisy measurement at time $t_i$ is modeled as
	\begin{equation}
		\widetilde{\bm{y}}_i = \exp([\bm{\theta}_i]_{\times}) \bm{y}_i,
	\end{equation}
	where $[\cdot]_{\times}$ denotes the skew-symmetric cross-product operator, and $\bm{\theta}_i \in \mathbb{R}^3$ is a zero-mean isotropic Gaussian rotation vector, $\bm{\theta}_i \sim \mathcal{N}(\mathbf{0},\sigma_\theta^2\mathbf{I}_3)$.
\end{assumption}

Under Assumption~\ref{ass:los_noise}, a first-order expansion for $\|\bm{\theta}_i\| \ll 1$ gives $\delta\bm{y}_i\equiv\widetilde{\bm{y}}_i-\bm{y}_i=\bm{z}_i\approx[\bm{\theta}_i]_{\times}\bm{y}_i=\bm{\theta}_i\times\bm{y}_i$, which is orthogonal to the LOS direction by construction. The associated covariance matrix is therefore
\begin{equation}
	\label{eq:measurement_covariance}
	\bm{\Sigma}_{\theta,i} = \sigma_\theta^2 \left( \mathbf{I}_{3} - \bm{y}_i \bm{y}_i^\top \right).
\end{equation}

\vspace{1em}
\subsection{First-Order Covariance Propagation}
\label{subsec:error_propagation}
To quantify the uncertainty of the initial-state estimate in~\eqref{eq:state_reconstruction} at $t_{N-1}$, we compute the covariance $\mathbf{P}_0\equiv\mathbf{P}_{0|N-1}$ by linearizing the mapping from LOS perturbations to the state estimation error and propagating the measurement noise through a first-order approximation.

Assuming small measurement errors, the perturbation of the scaling vector is derived by linearizing the least-squares solution~\cite{GGL16}. The perturbation $\delta \bm{k}_N = \hat{\bm{k}}_N - \bm{k}_N$ is obtained via a first-order expansion of~\eqref{eq:LS problem}. Since $\mathbf{B}_N$ is independent of the LOS measurements (implying $\delta \mathbf{B}_N = \mathbf{0}$), the resulting first-order variation is:
\begin{equation}
	\delta \bm{k}_N \approx -(\mathbf{A}_N^\top \mathbf{A}_N)^{-1} \mathbf{A}_N^\top \, \delta \mathbf{A}_N \, \bm{k}_N.
\end{equation}

With $\delta\mathbf{A}_N=\widetilde{\mathbf{A}}_N-\mathbf{A}_N$, $\delta\mathbf{C}=\widetilde{\mathbf{C}}-\mathbf{C}$, and $\delta\bm{x}_0=\hat{\bm{x}}_{0|N-1}-\bm{x}(t_0)$, substituting this into the linearized state reconstruction equation, $\delta\bm{x}_0 \approx \delta\mathbf{C}\,\mathbf{J}\bm{k}_N + \mathbf{C}\mathbf{J}\delta\bm{k}_N$, yields:
\begin{equation}
	\label{eq:dx0_matrix_form}
	\delta \bm{x}_0 \approx
	\delta \mathbf{C}\,\mathbf{J}\bm{k}_N -
	\mathbf{C}\mathbf{J}(\mathbf{A}_N^\top \mathbf{A}_N)^{-1}
	\mathbf{A}_N^\top \delta \mathbf{A}_N\, \bm{k}_N.
\end{equation}

Since $\mathbf{A}_N$ and $\mathbf{C}$ depend linearly on the LOS measurements, their perturbations can be decomposed as:
\begin{equation*}
	\delta \mathbf{A}_N = \sum_{i=0}^{N-1} \mathbf{A}_{(i)} \otimes \delta\bm{y}_i, \quad
	\delta \mathbf{C} = \sum_{i=0}^{N-1} \mathbf{C}_{(i)} \otimes \delta\bm{y}_i,
\end{equation*}
Here, $\otimes$ denotes tensor contraction along the vector component dimension. $\mathbf{A}_{(i)} = \partial \mathbf{A}_N / \partial \bm{y}_i$ and $\mathbf{C}_{(i)} = \partial \mathbf{C} / \partial \bm{y}_i$ are third-order tensors, representing the derivatives of matrices with respect to vector components. The operations $\mathbf{A}_{(i)} \otimes \delta\bm{y}_i$ and $\mathbf{C}_{(i)} \otimes \delta\bm{y}_i$ signify contraction along the third dimension, yielding matrices with the same dimensions as $\mathbf{A}_N$ and $\mathbf{C}$, respectively.

A key property facilitating practical computation is the linearity of these tensor contractions. Given the specific structure of the tensors, the following commutative property holds:
$
(\mathbf{C}_{(i)} \otimes \delta\bm{y}_i) \mathbf{J} \bm{k}_N 
	= (\mathbf{C}_{(i)} \otimes \mathbf{J} \bm{k}_N) \, \delta\bm{y}_i,
$
where $\mathbf{C}_{(i)} \otimes \mathbf{J} \bm{k}_N$ denotes tensor contraction along the second dimension of $\mathbf{C}_{(i)}$ with the vector $\mathbf{J} \bm{k}_N$, yielding a matrix of size $6 \times 3$. Similarly, $\mathbf{A}_{(i)} \otimes \bm{k}_N$ denotes contraction along the second dimension of $\mathbf{A}_{(i)}$ with $\bm{k}_N$, yielding a matrix of size $3(N-2) \times 3$.
Therefore, the sensitivity matrix can be written as
\begin{equation}\label{eq:S_i}
	\bm{S}_i = \mathbf{C}_{(i)} \otimes \mathbf{J} \bm{k}_N - \mathbf{C}\mathbf{J}(\mathbf{A}_N^\top \mathbf{A}_N)^{-1} \mathbf{A}_N^\top (\mathbf{A}_{(i)} \otimes \bm{k}_N),
\end{equation}
where $\bm{S}_i$ maps $\delta\bm{y}_i$ to $\delta\bm{x}_0$. 

The explicit forms of these tensor-vector products are:
\begin{itemize}
	\item \textbf{Variation in $\bm{y}_0$}:
	\begin{align*}
		\mathbf{C}_{(0)} \otimes \mathbf{J} \bm{k}_N &=
		\begin{bmatrix}
			k_0 \mathbf{I}_3 \\
			-k_0 \bm{\Phi}_{rv}^{-1}(T)\bm{\Phi}_{rr}(T)
		\end{bmatrix},\\
		\mathbf{A}_{(0)} \otimes \bm{k}_N &=
		k_0\,\left[\bm{\Phi}_{rr}(mT)\right. \\
		& \left.-\bm{\Phi}_{rv}(mT)\bm{\Phi}_{rv}^{-1}(T)\bm{\Phi}_{rr}(T)\right]_{m=2}^{N-1}.
	\end{align*}
	
	\item \textbf{Variation in $\bm{y}_1$}:
	\begin{align*}
		\mathbf{C}_{(1)} \otimes \mathbf{J} \bm{k}_N &=
		\begin{bmatrix}
			\mathbf{0}_{3\times 3} \\
			k_1 \bm{\Phi}_{rv}^{-1}(T)
		\end{bmatrix},\\
		\mathbf{A}_{(1)} \otimes \bm{k}_N &=
		k_1\,\left[\bm{\Phi}_{rv}(mT)\bm{\Phi}_{rv}^{-1}(T)\right]_{m=2}^{N-1}.
	\end{align*}
	
	\item \textbf{Variation in $\bm{y}_i$ ($i \geq 2$)}:
	\begin{align*}
		\mathbf{C}_{(i)} \otimes \mathbf{J} \bm{k}_N &= \mathbf{0}_{6\times 3}, \\
		\mathbf{A}_{(i)} \otimes \bm{k}_N &= -k_i\,
		\begin{bmatrix}
		\mathbf{0}_{3(i-2)\times3}\\
		\mathbf{I}_3\\
		\mathbf{0}_{3(N-1-i)\times3}
		\end{bmatrix}.
	\end{align*}
\end{itemize}
Here $[\cdot]_{m=2}^{N-1}$ denotes vertical stacking of the $3\times3$ blocks. Thus each $\mathbf{A}_{(i)}\otimes\bm{k}_N$ has size $3(N-2)\times3$; zero-row blocks are omitted at the endpoints.

Under Assumption~\ref{ass:los_noise} and for mutually uncorrelated LOS perturbations satisfying $\mathbb{E}[\delta\bm{y}_i \, \delta\bm{y}_j^\top] = \mathbf{0}$ for $i \neq j$, the estimation error admits the first-order approximation
\begin{equation}
	\label{eq:dx0_sum}
	\delta\bm{x}_0 \approx \sum_{i=0}^{N-1} \bm{S}_i \, \delta \bm{y}_i,
\end{equation}
with the sensitivity matrix $\bm{S}_i$ defined in~\eqref{eq:S_i}.
Since the cross terms vanish for uncorrelated perturbations, the first-order covariance of the batch IROD estimate is
\begin{equation}
	\label{eq:P0_analytical}
	\mathbf{P}_0 = \mathbb{E}[\delta\bm{x}_0 \, \delta\bm{x}_0^\top] = \sum_{i=0}^{N-1}
	\bm{S}_i \, \bm{\Sigma}_{\theta,i} \, \bm{S}_i^\top.
\end{equation}

\begin{remark}
Equation~\eqref{eq:P0_analytical} separates the effect of geometry from the effect of sensor quality. The matrices $\bm{S}_i$ encode how informative the maneuvered LOS geometry is, while $\bm{\Sigma}_{\theta,i}$ captures the contribution of sensor noise.
In implementation, the analytical covariance is evaluated using the measured LOS vectors and estimated scale factors in place of their ideal values.
\end{remark}

The formulation in~\eqref{eq:P0_analytical} holds for any $N \ge 3$ for which $\mathbf{A}_N$ has full column rank. In the minimal case of $N=3$, provided that condition~\eqref{eq:observability_condition} is satisfied, matrix $\mathbf{A}_N$ becomes square and invertible. Consequently, $(\mathbf{A}_N^\top \mathbf{A}_N)^{-1}\mathbf{A}_N^\top = \mathbf{A}_N^{-1}$, and the sensitivity matrices reduce to $\bm{S}_i = \mathbf{C}_{(i)} \otimes \mathbf{J} \bm{k}_N - \mathbf{C}\mathbf{J}\mathbf{A}_N^{-1}(\mathbf{A}_{(i)} \otimes \bm{k}_N)$, which is consistent with the square-system case.

\subsection{A Criterion to Quantify Observability}
\label{subsec:quantifying_observability}
The binary observability conditions in Sec.~\ref{subsec:observability_analysis} establish uniqueness of the initial-state reconstruction. Its practical accuracy also depends on measurement noise and conditioning. We therefore compare input-design strategies using the analytical covariance $\mathbf{P}_0$ and its condition number.

Since position and velocity have different units, we normalize the covariance using the orbital scales $a$ and $V=na$~\cite{HB83}, where $n=\sqrt{\mu/a^3}$ is the target's mean motion. With $\mathbf{T}=\mathrm{diag}(a^{-1},a^{-1},a^{-1},V^{-1},V^{-1},V^{-1})$, the normalized covariance is $\mathbf{P}_{0,\mathrm{norm}}=\mathbf{T}\mathbf{P}_0\mathbf{T}^\top$.

For symmetric positive-definite $\mathbf{P}_{0,\mathrm{norm}}$, the singular value decomposition (SVD) coincides with its eigendecomposition~\cite{Abd07}: $\mathbf{P}_{0,\mathrm{norm}}=\mathbf{U}\mathbf{\Sigma}\mathbf{U}^\top$. The orthogonal columns of $\mathbf{U}$ define the principal uncertainty axes and $\mathbf{\Sigma}=\mathrm{diag}(\sigma_1,\ldots,\sigma_6)$, with $\sigma_1\ge\cdots\ge\sigma_6>0$.

The singular values are the variances along these axes, with ellipsoid semi-axis lengths proportional to $\sqrt{\sigma_i}$. Their disparity is captured by the condition number
\begin{equation}
	 \kappa(\mathbf{P}_{0,\mathrm{norm}})=\frac{\sigma_{\max}}{\sigma_{\min}}=\frac{\sigma_1}{\sigma_6}.
\end{equation}
A large $\kappa$ means that some normalized state directions are estimated much less accurately than others, producing an anisotropic uncertainty ellipsoid. Conversely, $\kappa\approx1$ corresponds to comparable uncertainty across directions. A lower condition number therefore indicates a more balanced reconstruction.

The condition number describes relative uncertainty across directions and does not determine its absolute magnitude. We therefore interpret it together with the maximum eigenvalues of the position and velocity covariance blocks, which quantify their worst-case uncertainties.

\subsection{Handover to Rendezvous}
\label{subsec:handover}

We assess handover readiness using separate thresholds for position
and velocity uncertainty. Partition the batch covariance as
$\mathbf{P}_0 =
\bigl[\mathbf{P}_{rr}\;\mathbf{P}_{rv};
      \mathbf{P}_{vr}\;\mathbf{P}_{vv}\bigr]$.

\begin{proposition}
\label{prop:transition}
Let $\mathbf{P}_{rr}$ and $\mathbf{P}_{vv}$ denote the position and velocity blocks of the batch covariance matrix $\mathbf{P}_0$. The batch estimate is accepted for handover to the recursive filter if the worst-case position and velocity uncertainties satisfy
\begin{equation} \label{eq:trans_cond}
	\lambda_{\max}(\mathbf{P}_{rr}) < \sigma^2_{\text{pos}} \quad \text{and} \quad \lambda_{\max}(\mathbf{P}_{vv}) < \sigma^2_{\text{vel}}.
\end{equation}
\end{proposition}

Once condition~\eqref{eq:trans_cond} is satisfied, the estimate is propagated from the initial batch epoch $t_0$ to the current time step $t_i$ to initialize the sequential filter. The state is propagated using the linear dynamics from~\eqref{eq:CW_r_v}, and the covariance is updated as
\begin{equation}
	\label{cov. prop}
	\mathbf{P}(t_i) = \bm{\Phi}(t_i - t_0)\mathbf{P}_0 \bm{\Phi}(t_i - t_0)^\top + \mathbf{Q}_{\text{int}}(t_i - t_0),
\end{equation}
where the integrated process noise covariance $\mathbf{Q}_{\text{int}}(\ell) = \int_0^{\ell} \bm{\Phi}(\ell - \xi) \mathbf{Q} \bm{\Phi}(\ell - \xi)^\top d\xi$ accounts for unmodeled accelerations and disturbances, including the mismatch between the linearized CW dynamics and the true nonlinear motion, while $\mathbf{Q}$ is a user-defined continuous-time process noise intensity matrix.
In practice, this integral is approximated using a Riemann sum. 

At the handover time $t_{i_h}$, the propagated pair initializes the EKF as
$(\hat{\bm{x}}_{0|0}^{\mathrm{EKF}},\mathbf{P}_{0|0}^{\mathrm{EKF}})
=(\hat{\bm{x}}(t_{i_h}),\mathbf{P}(t_{i_h}))$.

\begin{remark}
Proposition~\ref{prop:transition} plays a different role from Proposition~\ref{prop:observability}. Observability ensures uniqueness of the reconstructed initial state, whereas the transition rule enforces an application-dependent accuracy requirement before the EKF is trusted to take over.
\end{remark}

\section{Active Learning for Input Design}\label{sec:IROD-OB}

To overcome the observability limitations identified in Sec.~\ref{subsec:observability_analysis}, we propose a dual-control framework enhanced by active learning. 
This framework synthesizes an excitation input sequence $\UU^* = \{\bm{u}_i^*\}_{i=0}^{N_u-1}$ designed to maximize the observability of the initial state $\bm{x}_0 = [\bm{r}_0^\top \ \bm{v}_0^\top]^\top$. 
As in Sec.~\ref{sec:IROD}, $\bm{u}_0=\bm{0}$, and $N_u$ input entries yield $N_u+1$ observations over the complete design horizon.
Using the continuous-time model~\eqref{eq:CW_equations_state_space} and the measurement model~\eqref{eq:LOS}, the discrete-time relative-motion dynamics are described by:
\begin{equation}
	\label{eq:LVLH}
    \begin{aligned}
		\bm{x}_{i+1} &= \mathbf{A}_T \bm{x}_{i} + \mathbf{B}_T \bm{u}_i, \quad i \in \Ni{0}{N_u-1} \\
        \widetilde{\bm{y}}_i &= \frac{\left[\mathbf{I}_{3} \ \ \mathbf{0}_{3}\right] \bm{x}_i}{\left\lVert\left[\mathbf{I}_{3} \ \ \mathbf{0}_{3}\right] \bm{x}_i\right\rVert_2} + \bm{z}_i, \quad i \in \Ni{0}{N_u} 
    \end{aligned}
\end{equation}
Here $\bm{x}_i=\bm{x}(t_i^-)$ on the same grid $t_i=t_0+iT$
used for batch IROD, with $\bm{u}_i$ applied after the measurement and
$\bm{z}_i=\delta\bm{y}_i$ as defined in Sec.~\ref{subsec:error model}.
The CW transition matrices are $\mathbf{A}_T=\bm{\Phi}(T)$ and
$\mathbf{B}_T=\bm{\Phi}(T)\mathbf{B}$.

For numerical propagation with a smaller step $\Delta t$, the
semi-implicit Euler matrices are
\begin{equation}\label{eq:CW_equations_state_space_discrete}
\begin{aligned}
\mathbf{A}_{\Delta} & =
\begin{bmatrix}
\mathbf{I}_3+\Delta t^2\mathbf{A}_{vr} &
\Delta t(\mathbf{I}_3+\Delta t\mathbf{A}_{vv})\\
\Delta t\mathbf{A}_{vr} & \mathbf{I}_3+\Delta t\mathbf{A}_{vv}
\end{bmatrix},\\
\mathbf{B}_{\Delta} & =
\begin{bmatrix}\Delta t\mathbf{I}_3\\\mathbf{I}_3\end{bmatrix},
\end{aligned}
\end{equation}
where $\mathbf{A}_{vr}$ and $\mathbf{A}_{vv}$ are the lower blocks of
$\mathbf{A}$. Over $q=T/\Delta t$ integration steps, with integer $q$,
$\mathbf{A}_T\approx\mathbf{A}_{\Delta}^{q}$ and
$\mathbf{B}_T\approx\mathbf{A}_{\Delta}^{q-1}\mathbf{B}_{\Delta}$,
with one impulse at the start of the interval.

Given a candidate input sequence $\UU = \{\bm{u}_i\}_{i=0}^{N_u-1}$ and the resulting set of LOS measurements $\{\widetilde{\bm{y}}_i\}_{i=0}^{N_u}$, the initial-state estimation problem is formulated as follows:
\begin{subequations} \label{eq:IROD_problem}
\begin{align}
    \hat{\bm{x}}^*_0 = \arg\min_{\hat{\bm{x}}_0} \ & \frac{1}{N_u+1} \sum_{i=0}^{N_u} \left\lVert\widetilde{\bm{y}}_i - \hat{\bm{y}}_i \right\rVert_2^2 + \lambda \left\lVert\hat{\bm{x}}_0\right\rVert_2^2\\
    \text{s.t.}\;\; 
	& \hat{\bm{y}}_i = \frac{\left[\mathbf{I}_{3} \ \ \mathbf{0}_{3}\right] \hat{\bm{x}}_i}{\left\lVert \left[\mathbf{I}_{3} \ \ \mathbf{0}_{3}\right] \hat{\bm{x}}_i \right\rVert_2}, \quad i \in \Ni{0}{N_u} \label{eq:measurement_model} \\
	& \hat{\bm{x}}_{i+1} = \mathbf{A}_T \hat{\bm{x}}_i + \mathbf{B}_T \bm{u}_i, \quad i \in \Ni{0}{N_u-1} \label{eq:dynamics_constraint} \\
	& \hat{\bm{x}}_0 = \hat{\bm{x}}_{0|N_u}, \label{eq:initial_condition}
\end{align}
\end{subequations}
where $\hat{\bm{y}}_i$ denotes the predicted measurement, $\hat{\bm{x}}_i$ the predicted state, $\hat{\bm{x}}_{0|N_u}$ the initial state estimate, and $\lambda \ge 0$ is a regularization parameter. 

The objective of the input design is to find the sequence $\UU^*$ that optimizes an observability criterion by maximizing exploration of the output space, thereby improving the estimation quality achieved by solving~\eqref{eq:IROD_problem}. The injection of known control inputs $\UU$ can satisfy the observability condition~\eqref{eq:observability_condition} by providing the necessary absolute scale reference. Since the future measurements $\widetilde{\bm{y}}_i$ are unavailable during the offline design phase, we treat the initial state $\bm{x}_0$ and the measurement noise as random variables and propose the following AL approaches.

\subsection{Expected State-Estimation Error Minimization}\label{sec:IROD-AL-offline-x0}
Given a nominal initial state $\bm{x}_0^{\text{nom}}$ (e.g., a preliminary estimate derived from initial observations or orbital mechanics), a noise sequence $\ZZ = \{\bm{z}_i\}_{i=0}^{N_u}$, and an input sequence $\UU$, we define the optimal state estimate as
\begin{equation}
	\hat{\bm{x}}_{0}^*(\bm{x}_0, \ZZ, \UU) = \arg \min_{\hat{\bm{x}}_0} \frac{1}{N_u+1} \sum_{i=0}^{N_u} \left\lVert \widetilde{\bm{y}}_i - \hat{\bm{y}}_i \right\rVert_2^2 + \lambda \left\lVert\hat{\bm{x}}_0\right\rVert_2^2,
	\label{eq:x0-est-A}
\end{equation}
where the input sequence $\UU$ is designed to minimize the expected estimation error relative to the nominal state:
\begin{align}
		\UU^* = & \arg\min_{\UU} \ \mathbb{E}_{\bm{x}_0^{\text{nom}}, \ZZ, \XX^{\text{ref}}} \left[
		\phantom{\sum_{i=0}^{N_u-1}}\hspace*{-1.5em}
		\left\lVert\hat{\bm{x}}_{0}^*(\bm{x}_0^{\text{nom}}, \ZZ, \UU) - \bm{x}_0^{\text{nom}}\right\rVert_2^2 \right. \notag \\
		& \left.+ \frac{\rho}{N_u} \sum_{i=1}^{N_u} \left\lVert\bm{x}_i - \bm{x}_i^{{\text{ref}}}\right\rVert_{\mathbf{Q}_x}^2 \right]
		+ \frac{\tau}{N_u} \sum_{i=0}^{N_u-1} \left\lVert\bm{u}_i\right\rVert_2^2.
	\label{eq:U-AL-offline-E1}
\end{align}
In~\eqref{eq:U-AL-offline-E1}, $\tau \geq 0$ is the regularization parameter penalizing the control effort, $\rho \geq 0$ weights the deviation from a reference trajectory $\{\bm{x}_i^{{\text{ref}}}\}_{i=1}^{N_u}$ (e.g., setting $\bm{x}_i^{\text{ref}} = \bm{x}_0^{\text{nom}}$ for station-keeping), and $\mathbf{Q}_x$ is a positive-definite state-tracking weighting matrix.

To render problem~\eqref{eq:U-AL-offline-E1} computationally tractable, we approximate the expectation using $M$ samples 
$\{\bm{x}_0^{s,\text{nom}}, \ZZ^s, \XX^{s,{\text{ref}}}\}_{s=1}^{M}$, resulting in the following bilevel optimization problem:
\begin{equation}
	\begin{aligned}
		& \UU^* = \arg\min_{\UU} \ \frac{1}{M} \sum_{s=1}^{M}  \left\lVert\hat{\bm{x}}_{0}^*(\bm{x}_0^{s, \text{nom}}, \ZZ^s, \UU) - \bm{x}_0^{s, \text{nom}}\right\rVert_2^2 \\
		& + \frac{\rho}{MN_u} \sum_{s=1}^{M} \sum_{i=1}^{N_u} \left\lVert\bm{x}_i^s - \bm{x}_i^{s,{\text{ref}}}\right\rVert_{\mathbf{Q}_x}^2
		+ \frac{\tau}{N_u} \sum_{i=0}^{N_u-1} \left\lVert\bm{u}_i\right\rVert_2^2,
		\label{eq:U-AL-offline-SUM1}
	\end{aligned}
\end{equation}
where $\hat{\bm{x}}_{0}^*(\bm{x}_0^{s, \text{nom}}, \ZZ^s, \UU)$ is the solution to~\eqref{eq:x0-est-A} for the specific realization $\bm{x}_0=\bm{x}_0^{s, \text{nom}}$ and $\ZZ=\ZZ^s$, and $\bm{x}_i^s$ denotes the state trajectory corresponding to the initial condition $\bm{x}_0^{s, \text{nom}}$ and input sequence $\UU$.

\subsection{Batch Active Learning (Greedy-\texorpdfstring{$y$}{y})}\label{sec:IROD-AL-offline-y}
An alternative approach to enhancing initial-state observability is to maximize the exploration of the output space. This method, termed Greedy-$y$, seeks to disperse the measurements by minimizing the following objective:
\begin{equation}
	\begin{aligned}
	\UU^* = & \arg\min_{\UU} \  \mathbb{E}_{\bm{x}_0^{\text{nom}}, \ZZ, \XX^{\text{ref}}} \left[
	- \frac{\gamma}{N_u+1} \sum_{i=0}^{N_u} d_y(\widetilde{\bm{y}}_i,\YY) \right. \\
	& \left.+ \frac{\rho}{N_u} \sum_{i=1}^{N_u} \left\lVert\bm{x}_i - \bm{x}_i^{{\text{ref}}}\right\rVert_{\mathbf{Q}_x}^2\right]
	+ \frac{\tau}{N_u} \sum_{i=0}^{N_u-1} \left\lVert\bm{u}_i\right\rVert_2^2,
	\label{eq:U-AL-offline-E2}
	\end{aligned}
\end{equation}
where $d_y(\widetilde{\bm{y}}_i,\YY) = \min_{h=0, h \neq i}^{N_u} \left\lVert\widetilde{\bm{y}}_i - \widetilde{\bm{y}}_h\right\rVert_2^2$ is the minimum squared distance in the output space between the output $\widetilde{\bm{y}}_i$ and the remaining samples in $\YY = \{\widetilde{\bm{y}}_h\}_{h=0}^{N_u}$, and $\gamma$ is a hyperparameter that balances the exploration of the output space with the fidelity of the state trajectory. The idea behind~\eqref{eq:U-AL-offline-E2} is to select the input sequence $\UU$ that disperses the output measurements $\widetilde{\bm{y}}_i$ as broadly as possible, similar to the AL methods for static regression proposed in~\cite{WLH19}.

Rather than maximizing the minimum distance $d_y(\widetilde{\bm{y}}_i,\YY)$, which is a non-differentiable objective, we minimize the sum of inverse squared distances to facilitate gradient-based optimization:
\begin{equation}
    \hat d_y(\widetilde{\bm{y}}_i,\YY)=\frac{1}{N_u} \sum_{\substack{h=0 \\ h \neq i}}^{N_u} \frac{1}{\left\lVert\widetilde{\bm{y}}_i - \widetilde{\bm{y}}_h\right\rVert_2^2 + \epsilon},
\end{equation}
where $\epsilon$ is a small positive constant to avoid division by zero. 
Replacing the expectations with empirical means over $M$ samples $\bm{x}_0^s$, $\ZZ^s$, and $\XX^{s,{\text{ref}}}$, $s \in \Ni{1}{M}$, leads to the following optimization problem:
\begin{align}
	& \UU^* = \arg\min_{\UU} \frac{\gamma}{MN_u(N_u+1)} \sum_{s=1}^{M} \sum_{i=0}^{N_u} \sum_{\substack{h=0 \\ h \neq i}}^{N_u}\frac{1}{\left\lVert\widetilde{\bm{y}}_i^s - \widetilde{\bm{y}}_h^s\right\rVert_2^2 + \epsilon} \notag \\
	& + \frac{\rho}{MN_u} \sum_{s=1}^{M} \sum_{i=1}^{N_u} \left\lVert\bm{x}_i^s - \bm{x}_i^{s,{\text{ref}}}\right\rVert_{\mathbf{Q}_x}^2
	+ \frac{\tau}{N_u} \sum_{i=0}^{N_u-1} \left\lVert\bm{u}_i\right\rVert_2^2,
	\label{eq:U-AL-offline-SUM2}
\end{align}
where $\widetilde{\bm{y}}_i^s$ are the measurements generated via~\eqref{eq:LVLH} using the initial state $\bm{x}_0^s$, input sequence $\UU$, and measurement noise $\ZZ^s$, with $\bm{x}_i^s$ denoting the corresponding state trajectory.

It is important to note that while the optimization problem in~\eqref{eq:U-AL-offline-SUM2} remains nonconvex, it avoids the bilevel structure of~\eqref{eq:U-AL-offline-SUM1}, thereby offering improved computational efficiency. Furthermore, the penalty term on the control input $\bm{u}_i$ in~\eqref{eq:U-AL-offline-SUM2} can be replaced by or augmented with explicit constraints, such as $\bm{u}_i \in \Omega_u$, where $\Omega_u$ represents the set of admissible inputs. Typical examples include box constraints $\Omega_u = \{\bm{u} \in \mathbb{R}^{n_u}: \bm{u}^{\text{lower}} \leq \bm{u} \leq \bm{u}^{\text{upper}}\}$ or a discrete set $\Omega_u = \{\bm{u}^1, \ldots, \bm{u}^M\}$, as used in pseudorandom binary sequence excitation (e.g., $M=2$).

\subsection{Sequential Active Learning Extension}\label{sec:IROD-AL-sequential}
For completeness, we also introduce a sequential AL extension that can be used as an online fallback if the covariance-based handover condition is not satisfied after the pre-designed input sequence. In this approach, at each time step $i$, the next control input is selected to maximize the exploration of the output space based on the current state estimate $\hat{\bm{x}}_{i|i}$ and the history of measurements $\{\widetilde{\bm{y}}_j\}_{j=0}^{i}$, subject to mission constraints:
\begin{equation}\label{eq:uk-AL-sequential-x}
	\begin{aligned}
		\bm{u}_i^* = & \arg\min_{\bm{u} \in \Omega_u} \frac{\gamma}{i+1} \sum_{j=0}^{i} \frac{1}{\left\lVert\hat{\bm{y}}_{i+1|i}(\bm{u})- \widetilde{\bm{y}}_j\right\rVert_2^2 + \epsilon} \\
		& + \rho \left\lVert\hat{\bm{x}}_{i+1|i}(\bm{u}) - \bm{x}_{i+1}^{\text{ref}}\right\rVert_{\mathbf{Q}_x}^2 + \tau \left\lVert\bm{u}\right\rVert_2^2,
	\end{aligned}
\end{equation}
where $\hat{\bm{x}}_{i+1|i}(\bm{u})$ denotes the predicted state at time $i+1$ given input $\bm{u}$, and $\hat{\bm{y}}_{i+1|i}(\bm{u})=h(\hat{\bm{x}}_{i+1|i}(\bm{u}))$ denotes the corresponding predicted ideal LOS.
This sequential design is computationally lightweight and enables additional informative measurements to be collected before handover to the sequential navigation and control phase.

\section{Handover to Closed-Loop Rendezvous}\label{sec:EKF_MPC}
Once the handover condition~\eqref{eq:trans_cond} is satisfied, the batch IROD estimate and its covariance are propagated to the handover time as described in Sec.~\ref{subsec:handover}.
The resulting pair that we denote as $(\hat{\bm{x}}_{0|0}^{\mathrm{EKF}}, \mathbf{P}_{0|0}^{\mathrm{EKF}})$ initializes the EKF--MPC architecture described below.

\subsection{Sequential State Estimation}
\label{subsec:EKF}

An EKF provides recursive state estimates for feedback control,
accounting for the nonlinear LOS measurement function
$h(\bm{x})=\bm{r}/\|\bm{r}\|_2$ under the linear CW dynamics.
During rendezvous, step $k$ corresponds to $t_{i_h}+k\Delta t$,
with $k=0$ at handover.
Using the discrete-time CW matrices evaluated at the rendezvous
sampling interval $\Delta t$, the prediction step is
\begin{equation}
	\begin{aligned}
	\hat{\bm{x}}_{k|k-1}
	&=\mathbf{A}_{\Delta}\hat{\bm{x}}_{k-1|k-1}+\mathbf{B}_{\Delta}\bm{u}_{k-1},\\
	\mathbf{P}_{k|k-1}
	&=\mathbf{A}_{\Delta}\mathbf{P}_{k-1|k-1}\mathbf{A}_{\Delta}^{\top}+\mathbf{Q}_d,
	\end{aligned}
	\label{eq:ekf_prediction}
\end{equation}
where the dynamics Jacobian is simply $\mathbf{F}=\mathbf{A}_{\Delta}$
and $\mathbf{Q}_d=\mathbf{Q}_{\mathrm{int}}(\Delta t)$ is the
process noise covariance. With
$\hat{\bm{y}}_{k|k-1}=h(\hat{\bm{x}}_{k|k-1})$, the measurement
Jacobian is $\mathbf{H}_k=
\bigl[(\mathbf{I}_3-\hat{\bm{y}}_{k|k-1}\hat{\bm{y}}_{k|k-1}^{\top})/
\|\hat{\bm{r}}_{k|k-1}\|_2\quad\mathbf{0}_3\bigr]$.
The measurement update is
\begin{equation}
	\begin{aligned}
	\mathbf{L}_k
	&=\mathbf{P}_{k|k-1}\mathbf{H}_k^{\top}
	\bigl(\mathbf{H}_k\mathbf{P}_{k|k-1}\mathbf{H}_k^{\top}+\mathbf{R}\bigr)^{-1},\\
	\hat{\bm{x}}_{k|k}
	&=\hat{\bm{x}}_{k|k-1}
	+\mathbf{L}_k(\widetilde{\bm{y}}_k-\hat{\bm{y}}_{k|k-1}),\\
	\mathbf{P}_{k|k}
	&=(\mathbf{I}_6-\mathbf{L}_k\mathbf{H}_k)\mathbf{P}_{k|k-1},
	\end{aligned}
	\label{eq:ekf_state_update}
\end{equation}
where $\mathbf{R}$ is the measurement noise covariance.
The corrected estimate $\hat{\bm{x}}_{k|k}$ is supplied to the MPC.

\subsection{Model Predictive Control}
\label{subsec:mpc}

A constrained MPC balances rendezvous tracking accuracy and control effort~\cite{RH03,DPK12,SBBHMRTT13}. At each step $k$, it predicts the motion over $H$ steps using the discrete-time CW model~\eqref{eq:CW_equations_state_space_discrete}, initialized with $\hat{\bm{x}}_{k|k}$.

Let $\hat{\bm{r}}_{k+j|k}$ and $\hat{\bm{v}}_{k+j|k}$ denote the predicted position and velocity at step $k+j$, computed at step $k$ under the candidate input sequence $\{\bm{u}_{k+j|k}\}_{j=0}^{H-1}$. For the prescribed
reference $(\bm{r}^{\mathrm{ref}},\bm{v}^{\mathrm{ref}})$, define $\bm{e}^{r}_{k+j|k}=\hat{\bm{r}}_{k+j|k}-\bm{r}^{\mathrm{ref}}$ and $\bm{e}^{v}_{k+j|k}=\hat{\bm{v}}_{k+j|k}-\bm{v}^{\mathrm{ref}}$.
The controller minimizes
\begin{align}
		J_k
		={}& \sum_{j=0}^{H-1}
		\Biggl(
		\frac{w_{\mathrm{p}}}{s_r^2}
		\|\bm{e}^{r}_{k+j|k}\|_2^2
		+
		\frac{w_{\mathrm{v}}}{s_v^2}
		\|\bm{e}^{v}_{k+j|k}\|_2^2
		+
		\frac{w_u}{u_{\max}^2}
		\|\bm{u}_{k+j|k}\|_2^2
		\Biggr)\notag \\
		&+
		w_{\mathrm{t}}
		\left(
		\frac{\|\bm{e}^{r}_{k+H|k}\|_2^2}{s_r^2}
		+
		\frac{\|\bm{e}^{v}_{k+H|k}\|_2^2}{s_v^2}
		\right),
	\label{eq:mpc_cost}
\end{align}
subject to the prediction dynamics and the prescribed state and input constraints, including $\|\bm{u}_{k+j|k}\|_\infty\le u_{\max}$.
Here $s_r,s_v>0$ are position and velocity normalization scales, and $w_{\mathrm{p}},w_{\mathrm{v}},w_u,w_{\mathrm{t}}$ are nonnegative tuning weights.

Only the first optimal input, $\bm{u}_k=\bm{u}_{k|k}^{*}$, is applied,
and the optimization is repeated with the updated EKF estimate.
This closed-loop phase validates the proposed IROD initialization.
\begin{algorithm}[t!]
	\caption{AL-Enhanced IROD Framework with Downstream Rendezvous Validation}
	\label{alg:AL_IROD}
	\textbf{Input}: Offline horizon $N_{\text{off}}$ ($N_u=N_{\text{off}}$), thresholds $\sigma_{\text{pos}},\sigma_{\text{vel}}$, maximum number of steps $N_{\max}$, initial guess (optional).
	\vspace*{.05cm}\hrule\vspace*{.05cm}
	\begin{enumerate}[label*=\arabic*., ref=\theenumi]
		\item \textbf{Offline Phase}: Compute an optimal input sequence $\{\bm{u}^{\text{off}}_i\}_{i=0}^{N_{\text{off}}-1}$ via~\eqref{eq:U-AL-offline-SUM1} or~\eqref{eq:U-AL-offline-SUM2}.
		\item \textbf{IROD Phase}: Acquire $\widetilde{\bm{y}}_0$; for $i = 0, 1, \dots, N_{\max}-1$: \label{algo:AL_IROD:IROD}
		\begin{enumerate}[label=\theenumi.\arabic{enumii}]
			\item \textbf{Input Application}:
            \begin{itemize}
                \item If $i < N_{\text{off}}$: $\bm{u}_i \gets \bm{u}^{\text{off}}_i$.
                \item Else: $\bm{u}_i \gets \bm{0}$. The optional sequential design in~\eqref{eq:uk-AL-sequential-x} can replace this step when online re-excitation is desired.
            \end{itemize}
            \item Apply $\bm{u}_i$, propagate the dynamics, and collect the measurement $\widetilde{\bm{y}}_{i+1}$.
			\item \textbf{Estimation}: If $i+1 \ge 2$:
            \begin{itemize}
                \item Solve LS~\eqref{eq:LS problem} and compute~\eqref{eq:state_reconstruction} for $\hat{\bm{x}}_{0|i+1}$.
                \item Compute the analytical covariance $\mathbf{P}_{0|i+1}$ via~\eqref{eq:P0_analytical}.
            \end{itemize}
			\item \textbf{Transition Check}: If $i+1 \ge 2$:
            \begin{itemize}
                \item If the handover condition~\eqref{eq:trans_cond} is satisfied, set $i_h=i+1$:
				propagate $(\hat{\bm{x}}_{0|i+1}, \mathbf{P}_{0|i+1})$
				to $t_{i+1}$ using~\eqref{eq:CW_r_v} and~\eqref{cov. prop}
				to initialize the EKF; then proceed to the Rendezvous Phase
				(Step~\ref{algo:AL_IROD:rendezvous}).
            \end{itemize}
		\end{enumerate}
		\item \textbf{Rendezvous Phase} (Recursive Loop, with $k=0$ at handover): \label{algo:AL_IROD:rendezvous}
		\begin{enumerate}[label=\theenumi.\arabic{enumii}]
			\item \textbf{EKF}: At $k=0$, use the handover estimate and covariance. For $k\ge1$, predict the state and covariance~\eqref{eq:ekf_prediction} and update them using the new measurement $\widetilde{\bm{y}}_k$~\eqref{eq:ekf_state_update}.
            \item \textbf{MPC}: Compute the constrained rendezvous input from the current EKF estimate and apply it~\eqref{eq:mpc_cost}.
		\end{enumerate}
	\end{enumerate}
	\vspace*{.05cm}\hrule\vspace*{.05cm}
	\textbf{Output}: Trajectory to the target.
\end{algorithm}
The overall algorithm integrating these approaches is summarized in Algorithm~\ref{alg:AL_IROD}.

\section{Numerical Experiments}\label{sec:num-exp}

We assess the performance of Algorithm~\ref{alg:AL_IROD} (\ALIROD) in addressing the angles-only IROD problem. The numerical results focus on the batch formulation for offline input design in~\eqref{eq:U-AL-offline-E2}, which constitutes the main input-design strategy considered in this paper. The optional sequential AL formulation in Sec.~\ref{sec:IROD-AL-sequential} is included as an online extension and fallback mechanism for cases in which the covariance-based handover condition is not satisfied after the pre-designed input sequence. For numerical validation, the target and chaser are propagated independently in the Earth-centered inertial (ECI) frame using nonlinear two-body (Keplerian) dynamics, starting from the respective inertial states corresponding to the same initial orbital configuration used for the CW model. Between control inputs, each inertial position $\bm{r}^{\mathrm{ECI}}$ satisfies $\ddot{\bm{r}}^{\mathrm{ECI}}=-\mu\bm{r}^{\mathrm{ECI}}/\|\bm{r}^{\mathrm{ECI}}\|_2^3$; the control inputs are applied to the chaser.
The propagated inertial states are then transformed into the target-centered LVLH frame to obtain the relative position and velocity and generate the LOS measurements. The estimator and controller use the linear CW model~\eqref{eq:CW_equations_state_space}, introducing model mismatch. The dedicated covariance-validation experiment uses CW propagation to isolate the contribution of LOS noise.

Both spacecraft are assumed to be in near-circular low Earth orbits, as this is the primary operational domain for proximity operations such as rendezvous. To establish a comparable benchmark, the target's initial orbital elements are adopted from a LEO test case in~\cite{GLL18}. The initial orbital elements are listed in Table~\ref{tab:reference_orbit}. The chaser's initial relative state in the LVLH frame is defined as $\bm{r}_0 = \left [x_0 \ 0 \ 0 \right ]^\top$, $\bm{v}_0 = \left [0 \ 0 \ 0 \right ]^\top$, where $x_0$ represents the initial along-track displacement in the +V-bar direction (aligned with the target's orbital velocity vector).
This V-bar station-keeping scenario represents a worst-case configuration for observability, as the natural dynamics of the unperturbed CW system tend to maintain constant relative position and velocity. The absence of inherent relative motion makes the observability problem particularly challenging~\cite{GP15}.

\begin{table}[t]
\centering
\caption{Initial target orbital elements~\cite{GLL18}.}
\label{tab:reference_orbit}
\normalsize
\renewcommand{\arraystretch}{1.0}
\setlength{\tabcolsep}{3pt}
\begin{tabular}{lrlr}
\toprule
Parameter & Value & Parameter & Value \\
\midrule
Semi-major axis & $6790.1\,\mathrm{km}$ & Eccentricity & $0.001$ \\
Inclination & $51.6455^\circ$ & RAAN & $281.6522^\circ$ \\
Arg. of perigee & $37.3945^\circ$ & True anomaly & $322.7645^\circ$ \\
\bottomrule
\end{tabular}
\end{table}

We set the maximum experiment horizon to $N_{\max}=10$, corresponding to $N_{\max}+1=11$ LOS observations. The nominal IROD measurement and maneuver interval is $T=600\,\mathrm{s}$; $\Delta t=1\,\mathrm{s}$ denotes the numerical propagation step and the EKF--MPC update interval during rendezvous. The noise in the camera measurements is modeled as described in Sec.~\ref{subsec:error model}. The rotation vector is drawn from $\bm{\theta}_i\sim\mathcal{N}(\mathbf{0},\sigma_\theta^2\mathbf{I}_3)$, with nominal standard deviation $\sigma_\theta=10^{-4}\,\mathrm{rad}$ for each component. 
The initial state, $\bm{x}_0$, is recursively estimated via least-squares~\eqref{eq:LS problem}, which requires at least three observations. 
All computations were performed in MATLAB R2024b on a machine with an Intel(R) Core(TM) i7-8750H CPU @ 2.20 GHz and 16 GB of RAM.

\begin{figure}[t]
\centering
\includegraphics[width=1.0\columnwidth]{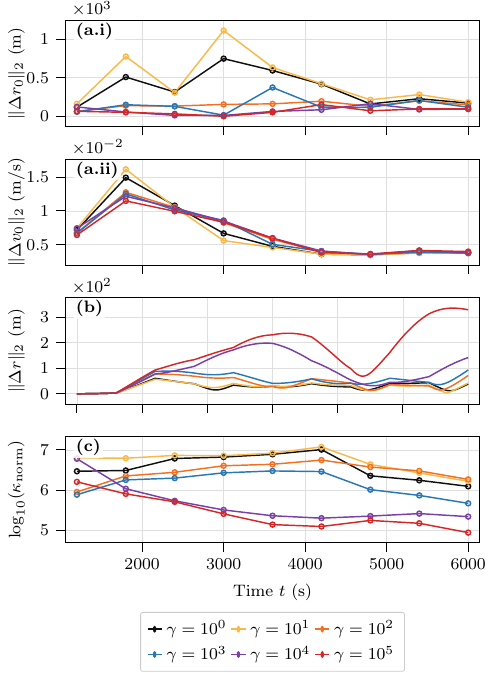}
\caption{Effect of the exploration weight $\gamma$ on the AL-IROD input design for $x_0 = 5000 \, \mathrm{m}$, with $\gamma$ ranging from $1$ to $10^5$: initial position and velocity estimation errors (a.i, a.ii), the station-keeping deviation (b), and the condition number of the normalized covariance matrix (c).}
	\label{fig:gamma_sweep_cmp}
\end{figure}

\subsection{Performance Analysis for V-Bar Station-Keeping}\label{sec:perf} 
In this experiment, we evaluate three offline input-design strategies based on the batch optimization formulation~\eqref{eq:U-AL-offline-E2}. The goal is to generate an input sequence that balances observability for state estimation with station-keeping performance for an angles-only IROD problem with a nominal initial relative distance of $\bar x_0 = 5000 \, \mathrm{m}$. The nominal initial state represents coarse prior information that may be available from ground-based observations or previous tracking. The uncertainty around this nominal state reflects the limited accuracy of such information. The IROD phase refines the initial-state estimate on board using angular measurements collected under known excitation maneuvers, providing the initialization for subsequent rendezvous navigation. The objective function in~\eqref{eq:U-AL-offline-E2} is nonconvex due to the exploration term. To handle the expectation over the initial-state uncertainty, we approximate the stochastic objective using a sample-average approach. Specifically, we select five representative initial states, $\bm{x}_0$, with along-track positions of $\{4800, \ 4900, \ 5000, \ 5100, \ 5200\} \, \mathrm{m}$ and zero initial relative velocity. The resulting optimization problem is solved using particle swarm optimization (PSO)~\cite{KE95}.

For all strategies, the control inputs are bounded by $\bm{u}^{\text{lower}} = -0.1 \, \mathrm{m/s} \cdot \bm{1}$ and $\bm{u}^{\text{upper}} = 0.1 \, \mathrm{m/s} \cdot \bm{1}$. The weights $\mathbf{Q}_x$ and $\rho$ apply to state errors expressed in km and km/s, and $\tau$ to impulses expressed in km/s; the reference state and control bounds are reported here in SI units ($0.1\,\mathrm{m/s}=10^{-4}\,\mathrm{km/s}$). The regularization parameters are set to $\rho = 10^8$ to strongly enforce station-keeping for $\bm{x}^{\text{ref}} = [5000,\, 0,\, 0,\, 0,\, 0,\, 0]^\top$, and $\tau = 10^{-2}$ to penalize control effort. Additionally, we enforce $\bm{u}_0 = \bm{0}$ in all experiments, optimizing the sequence $\{\bm{u}_i\}_{i=1}^{N_{\text{off}}-1}$ with $N_u=N_{\text{off}}=10$.

\begin{figure}[t]
\centering
\includegraphics[width=1.0\columnwidth]{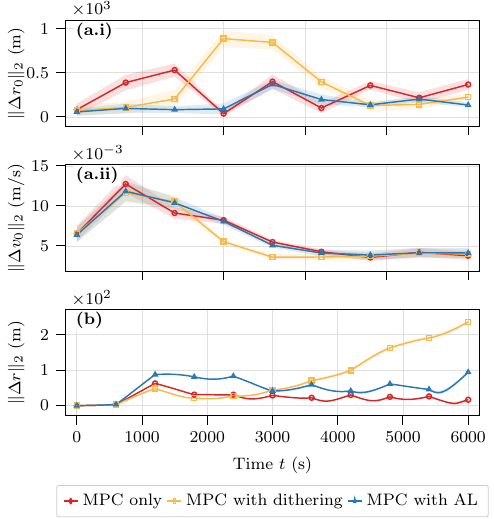}
\caption{Monte Carlo comparison of offline input design strategies for the open-loop case with initial relative distances in the range $x_0 \in [4750,5250] \, \mathrm{m}$. Panels (a.i) and (a.ii) show the initial position and velocity estimation errors; panel (b) shows the station-keeping deviation. The solid curves show the Monte Carlo mean, and the shaded areas represent one standard deviation.}
	\label{fig:mc_MPC_OL}
\end{figure}

We generate three distinct input sequences for comparison:

\noindent \textbf{MPC only}: This baseline strategy represents a purely control-oriented approach. It utilizes the same optimization framework but with the exploration term disabled ($\gamma = 0$). The state-tracking weight matrix is set to $\mathbf{Q}_x = \mathbf{I}_{6}$, treating all state deviations equally.

\noindent \textbf{MPC with dithering}: This strategy serves as a simple heuristic for exploration. First, an input sequence is generated using the MPC-only settings but with tighter control bounds of $\pm 0.09 \, \mathrm{m/s}$. A uniform random dither within $\pm 0.01 \, \mathrm{m/s}$ is then superimposed on this sequence to introduce excitation.

\noindent \textbf{MPC with AL}: This is the proposed AL approach, which directly optimizes~\eqref{eq:U-AL-offline-E2}. The state-tracking weight matrix is defined as $\mathbf{Q}_x = \text{diag}(250, 25, 25, 1, 1, 1)$, prioritizing stability in the along-track ($x$) direction while permitting exploratory maneuvers in the cross-track plane. We select $\gamma = 10^3$ based on an evaluation of the impact of the exploration weight $\gamma$ at an initial relative distance of $x_0 = 5000 \, \mathrm{m}$. The selection of $\gamma$ involves a trade-off: a larger $\gamma$ promotes better exploration, which lowers the initial-state estimation error and the covariance condition number, but it also induces larger deviations from the initial position, as shown in Fig.~\ref{fig:gamma_sweep_cmp}. The PSO algorithm is employed to solve~\eqref{eq:U-AL-offline-SUM2} with a swarm size of 100 and a maximum of 100 iterations. The computation time for each optimization run is approximately 2.5 hours.

\begin{figure}[t]
\centering
\includegraphics[width=1.0\columnwidth]{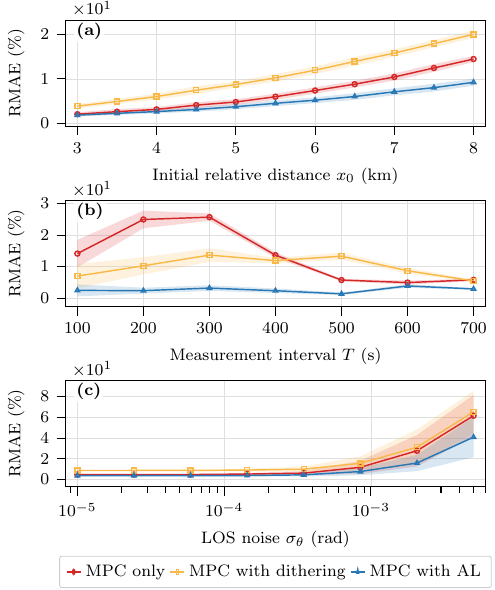}
\caption{Monte Carlo sensitivity analysis of the offline input design strategies using the RMAE metric over $M=100$ runs. The plots show sensitivity to the initial relative distance $x_0$ (Fig.~\ref{fig:sensitivity_rmae}(a)), measurement interval $T$ (Fig.~\ref{fig:sensitivity_rmae}(b)), and LOS noise level $\sigma_{\theta}$ (Fig.~\ref{fig:sensitivity_rmae}(c)). The solid curves show the Monte Carlo mean, and the shaded areas represent one standard deviation.}
    \label{fig:sensitivity_rmae}
\end{figure}

Subsequently, we validate these pre-designed input sequences in a Monte Carlo simulation with initial relative distances uniformly sampled from $x_0 \in [4750,5250] \, \mathrm{m}$ and independent measurement noise realizations, thereby testing robustness to mismatch between the design point and the realized initial condition. In Fig.~\ref{fig:mc_MPC_OL}, the solid curves show the Monte Carlo mean, and the shaded areas represent one standard deviation. The figure illustrates that the MPC-only design, while effective at station-keeping, fails to produce sufficiently informative maneuvers due to the absence of an exploration term, resulting in large estimation errors, especially at the beginning of the simulation. The MPC-with-dithering approach, despite introducing excitation, degrades both estimation and station-keeping performance, as the random dither is unstructured and does not account for the mission constraints. The MPC-with-AL strategy achieves superior estimation throughout the simulation while maintaining station-keeping, demonstrating that a principled joint optimization of observability and control performance is necessary to avoid the limitations of both baseline approaches.

\begin{figure*}[!t]
	\centering
	\includegraphics[width=1.0\textwidth]{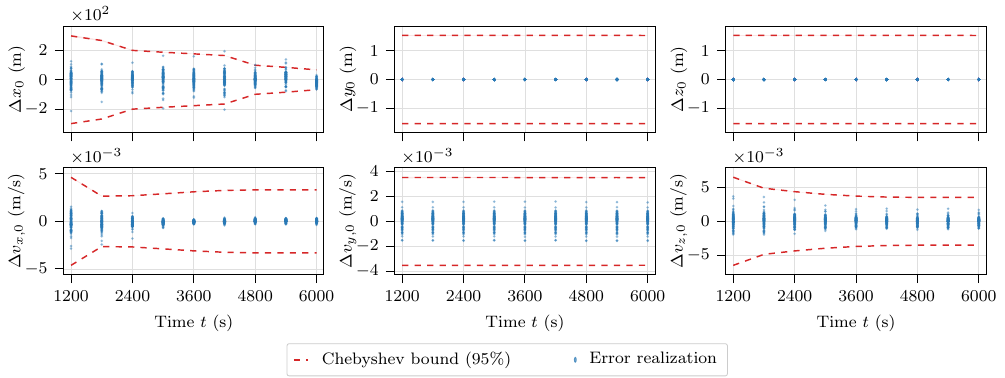}
	\caption{Monte Carlo validation ($M=100$ runs) of the analytical covariance for \ALIROD. The red dashed lines represent the averaged 95\% Chebyshev bounds derived from the diagonal elements of $\mathbf{P}_{0|i}$, while the blue dots represent the actual estimation errors $\hat{\bm{x}}_{0|i}-\bm{x}_0$ for position (upper) and velocity (lower).}
	\label{fig:cov_al}
\end{figure*}

\subsection{Sensitivity Analysis}\label{sec:sensitivity}
To assess the robustness of the offline input design strategies, we perform Monte Carlo sensitivity analyses with respect to three operational parameters: the initial relative distance $x_0$, the measurement interval $T$, and the LOS noise standard deviation $\sigma_\theta$. For each parameter value and each strategy, we run $M=100$ simulations with independent measurement noise realizations. Performance is quantified using the relative mean absolute error (RMAE) of the initial position estimate, normalized by the initial relative distance $x_0$:
\begin{equation}
	\text{RMAE} = \frac{1}{N_{\max}-i_{\min}+1} \sum_{i=i_{\min}}^{N_{\max}} \frac{\left\lVert\hat{\bm{r}}_{0|i}-\bm{r}_0\right\rVert_2}{x_0} \times 100\%.
	\label{eq:RMAE}
\end{equation}
We set $i_{\min}=5$, corresponding to the fourth available batch estimate, to exclude the initial transient. For the initial-distance, measurement-interval, and LOS-noise analyses, the final measurement index is fixed at $N_{\max}=10$, so the RMAE is averaged over estimates indexed from $i=5$ through $i=10$. 

\subsubsection{Initial Relative Distance}
Figure~\ref{fig:sensitivity_rmae}(a) reports the RMAE for initial relative distances from $3$ km to $8$ km with $0.5$ km increments, while the remaining parameters are fixed at their nominal values. The input sequence used in this test is the one designed for the nominal case and is not re-optimized for each initial relative distance. As the initial relative distance increases, the same bounded control inputs produce smaller LOS variations, reducing the relative information content of the maneuver sequence for the IROD reconstruction. Consequently, the RMAE tends to increase for all strategies, and larger excitation inputs or distance-dependent input redesign may be required at larger separations. Nevertheless, the MPC-with-AL strategy consistently achieves the lowest error over the tested range, showing that the optimized excitation remains more informative than the baseline inputs even away from the nominal design condition.

\subsubsection{Measurement Interval}
Figure~\ref{fig:sensitivity_rmae}(b) evaluates measurement intervals from $100$ s to $700$ s. Short intervals provide limited dynamic evolution between LOS measurements, which can degrade observability~\cite{GP15}. This effect is particularly evident for the MPC-only and MPC-with-dithering strategies, whose errors increase significantly at small $T$. In contrast, the AL-based design remains bounded and consistently more accurate across the tested sampling intervals, indicating that the designed excitation injects informative geometry already in the early part of the observation window.

\subsubsection{LOS Noise Level}
Figure~\ref{fig:sensitivity_rmae}(c) shows the sensitivity to LOS noise levels in the range $\sigma_\theta \in [10^{-5},5\times10^{-3}]$ rad. As expected, estimation accuracy degrades with increasing measurement noise, while the MPC-with-AL strategy preserves a clear advantage over the baseline strategies.

\begin{figure}[ht]
	\centering
	\includegraphics[width=1.0\columnwidth]{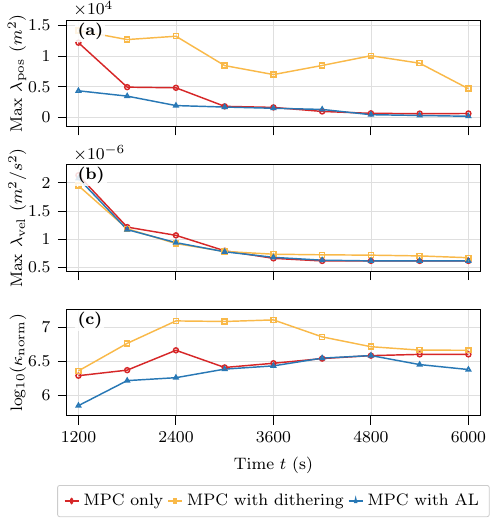}
	\caption{Comparison of covariance evolution metrics: maximum eigenvalue of position covariance (top), maximum eigenvalue of velocity covariance (middle), and condition number of the normalized covariance matrix (bottom).}
	\label{fig:cov_comparison}
\end{figure}

\subsection{Covariance Analysis: Validation and Comparison}\label{sec:cov_analysis}
To validate the analytical covariance matrix $\mathbf{P}_0$ derived in~\eqref{eq:P0_analytical}, we performed a Monte Carlo analysis ($M=100$ runs) using the linearized CW dynamics with the \ALIROD~strategy. This approach ensures consistency with the derivation assumptions, isolating the effects of measurement noise propagation from linearization errors. The initial relative state was fixed at $x_0 = 4850 \, \mathrm{m}$ to introduce a mismatch relative to the nominal value $\bar{x}_0 = 5000 \, \mathrm{m}$, while measurement noise realizations were randomized. As shown in Fig.~\ref{fig:cov_al}, the red dashed lines represent the arithmetic mean of the 95\% confidence intervals computed via Chebyshev's inequality~\cite{BLK01} across the Monte Carlo runs. The averaged covariance-derived bounds accurately enclose the empirical error distribution for both position and velocity components and shrink as observations accumulate.

We further assess the efficacy of the AL approach by comparing its covariance evolution with those of the other two strategies. Figure~\ref{fig:cov_comparison} displays the maximum eigenvalues of the position and velocity covariance blocks, $\lambda_{\max}(\mathbf{P}_{rr})$ and $\lambda_{\max}(\mathbf{P}_{vv})$, and the condition number $\kappa(\mathbf{P}_{0,\text{norm}})$ over time. Although all methods exhibit bounded and generally non-increasing eigenvalue trends, the AL strategy provides a clear advantage in position estimation, with consistently lower eigenvalues that demonstrate richer information content in the LOS measurements. While the velocity covariance behavior is similar across all methods, the AL solution maintains the lowest condition numbers, indicating better-conditioned estimation with a more uniform uncertainty distribution across the state space and enhanced numerical robustness.

\begin{figure}[!t]
	\centering
	\includegraphics[width=1.0\columnwidth]{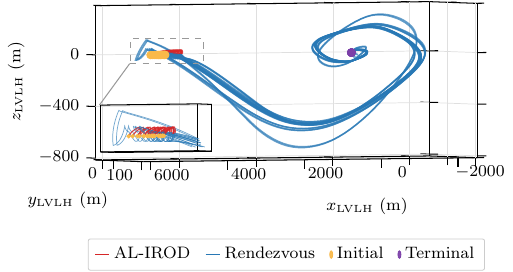}
	\caption{Hybrid \ALIROD~(red lines) and rendezvous trajectories (blue lines) in the LVLH frame for 20 initial positions (yellow dots) uniformly sampled from $x_0 \in [4750,5250]\,\mathrm{m}$ to the desired terminal position (purple dot) at $x_{\mathrm{LVLH}}=50\,\mathrm{m}$ relative to the target.}
	\label{fig:traj_MPC_EKF}
\end{figure}

\subsection{Closed-Loop Rendezvous Validation: IROD to Rendezvous}\label{sec:mpc_ekf}
We evaluate the complete pipeline in Algorithm~\ref{alg:AL_IROD} over 20 simulations, with $x_0$ sampled uniformly from $[4750,5250]\,\mathrm{m}$ and all remaining initial-state components set to zero. Handover uses the thresholds $\sigma_{\mathrm{pos}}^2=2000\,\mathrm{m}^2$ and $\sigma_{\mathrm{vel}}^2=0.005\,\mathrm{m}^2/\mathrm{s}^2$.
The propagated IROD estimate and covariance initialize the EKF--MPC rendezvous controller.

The continuous-time process noise intensity is set to $\mathbf{Q}=\mathrm{blkdiag}(10^{-10}\mathbf{I}_3,10^{-12}\mathbf{I}_3)$, as in~\eqref{cov. prop}, and the corresponding EKF process noise covariance is $\mathbf{Q}_d=\mathbf{Q}_{\mathrm{int}}(\Delta t)$. The measurement noise covariance used by the EKF is $\mathbf{R}=500\sigma_{\theta}^2\mathbf{I}_3$, with $\sigma_{\theta}=10^{-4}\,\mathrm{rad}$.

The rendezvous MPC uses the distance-dependent parameters in Table~\ref{tab:mpc_parameters}, with
$d=\|\hat{\bm{r}}_{k|k}\|_2$. At each optimization step, the normalization scales are $s_r=\max\{d,1000\,\mathrm{m}\}$ and $s_v=\max\{\|\hat{\bm{v}}_{k|k}\|_2,100\,\mathrm{m/s}\}$.
The input bound is $u_{\max}=0.5\,\mathrm{m/s}$ per component; the applied command is additionally limited to a velocity-increment norm of $0.1\,\mathrm{m/s}$ at $\Delta t=1\,\mathrm{s}$.

\begin{table}[t]
\centering
\caption{Distance-dependent rendezvous MPC parameters ($d$ in meters).}
\label{tab:mpc_parameters}
\normalsize
\renewcommand{\arraystretch}{1.12}
\setlength{\tabcolsep}{3pt}
\begin{tabular}{lccc}
\toprule
Parameter & $d>1000$ & $100<d\le1000$ & $d\le100$ \\
\midrule
$H$ & $10$ & $10$ & $8$ \\
$w_{\mathrm{p}}$ & $1$ & $2$ & $5$ \\
$w_{\mathrm{v}}$ & $2$ & $5$ & $10$ \\
$w_u$ & $10^{-1}$ & $10^{-1}$ & $10^{-6}$ \\
$w_{\mathrm{t}}$ & $10$ & $20$ & $50$ \\
Position bound [m] & $5000$ & $1000$ & $200$ \\
Velocity bound [m/s] & $2000$ & $1000$ & $500$ \\
\bottomrule
\end{tabular}
\end{table}
The state bounds in Table~\ref{tab:mpc_parameters} apply
componentwise to the predicted position and velocity.

Figure~\ref{fig:traj_MPC_EKF} shows the complete hybrid trajectories in the LVLH frame for 20 independent simulations. The yellow dots denote the sampled initial positions, while the purple dot denotes the desired final position, which is set to $x_{\mathrm{LVLH}}=50\,\mathrm{m}$ relative to the target. Each trajectory is composed of an \ALIROD~segment, shown in red, followed by the EKF--MPC rendezvous segment, shown in blue, after the covariance-based handover.

Across the 20 simulations, the proposed pipeline achieves successful convergence to the desired terminal region. The mean final position error is $1.388\,\mathrm{m}$, the mean final velocity error is $0.0514\,\mathrm{m/s}$, and the mean convergence time is $5389.1\,\mathrm{s}$. These results show that the IROD algorithm provides a sufficiently accurate prior for the EKF, allowing the MPC to complete the rendezvous despite variations in the initial position, measurement noise, and the mismatch between the nonlinear Keplerian propagation used in simulation and the CW model used by the estimator and controller.

\section{Conclusion}\label{sec:cons}

This paper presented an AL-based input-design framework for angles-only IROD. The approach integrates informative maneuver design, batch state reconstruction, analytical covariance estimation, and closed-loop rendezvous validation.

The primary practical benefit of the proposed framework is that it provides reliable initialization for subsequent autonomous proximity operations. By designing maneuvers that increase the information content of the LOS measurements, the method mitigates the scale ambiguity of angles-only measurements while respecting station-keeping constraints. The analytical covariance provides a quantitative assessment of the resulting initialization uncertainty and enables a covariance-based handover to sequential navigation.

Numerical simulations demonstrated that the AL design consistently improves estimation accuracy over baseline strategies and that the covariance-derived bounds reliably characterize the estimation errors. Successful closed-loop rendezvous maneuvers further validated the accuracy of the resulting IROD initialization.

\bibliographystyle{IEEEtran}
\bibliography{IROD_Input_Design_bib}

@STRING{INFSCI        = "Inf. Sci."}

@STRING{ICNN        = "Proc. Int. Conf. Neural Netw. ({ICNN})"}

@STRING{SLAIML        = "Synth. Lect. Artif. Intell. Mach. Learn."}

@STRING{CDC           = "Proc. {IEEE} Conf. Decis. Control ({CDC})"}

@STRING{IEEETAES      = "{IEEE} Trans. Aerosp. Electron. Syst."}

@STRING{IEEETCST      = "{IEEE} Trans. Control Syst. Technol."}

@STRING{ASTRODYN      = "Astrodynamics"}

@STRING{JGCD          = "J. Guid. Control Dyn."}

@STRING{JSEE          = "J. Syst. Eng. Electron."}

@STRING{ACTAASTRO     = "Acta Astronaut."}

@STRING{BITZSYM       = "Proc. Itzhack Y. Bar-Itzhack Meml. Symp. Estim. Navig. Spacecr. Control"}

@STRING{JSR           = "J. Spacecr. Rockets"}

@STRING{ENCMEASSTAT   = "Encycl. Meas. Stat."}

@STRING{JAS           = "J. Astronaut. Sci."}

@STRING{CURRCHINSCI   = "Curr. Chin. Sci."}

@STRING{SCITECH       = "AIAA SciTech Forum"}

@STRING{PFDGA         = "Prog. Flight Dyn. GNC Avionics"}

@STRING{IJRNC         = "Int. J. Robust Nonlinear Control"}

@STRING{AIAAGNC       = "AIAA Guid. Navig. Control Conf. Exhibit"}

@article{Bem23c,
    author={A. Bemporad},
    title={Active Learning for Regression by Inverse Distance Weighting},
    year=2023,
    journal=INFSCI,
    volume={626},
    pages={275--292},
    doi={10.1016/j.ins.2023.01.028},
    note={Code available at \url{http://cse.lab.imtlucca.it/~bemporad/ideal}},
}

@incollection{Set12,
  title={Active Learning},
  author={B. Settles},
  booktitle=SLAIML,
  number={18},
  publisher={Morgan \& Claypool Publishers},
  year={2012},
  doi={10.2200/S00429ED1V01Y201207AIM018},
}

@inproceedings{KE95,
  author={J. Kennedy and R. Eberhart},
  booktitle=ICNN,
  title={Particle Swarm Optimization}, 
  year={1995},
  volume={4},
  pages={1942-1948},
  doi={10.1109/ICNN.1995.488968}}

@article{WLH19,
  title={Active Learning for Regression Using Greedy Sampling},
  author={D. Wu and C. T. Lin and J. Huang},
  journal=INFSCI,
  volume={474},
  pages={90--105},
  year={2019},
  doi={10.1016/j.ins.2018.09.060},
}

@inproceedings{XB24,
    title = {Online Design of Experiments by Active Learning for System Identification of Autoregressive Models},
    booktitle = CDC,
    pages = {7202--7207},
    author = {K. Xie and A. Bemporad},
    year = {2024},
    doi={10.1109/cdc56724.2024.10886678},
}

@article{WG09,
    title = {Observability Criteria for Angles-Only Navigation},
    volume = {45},
    issn = {0018-9251},
    doi = {10.1109/TAES.2009.5259193},
    pages = {1194--1208},
    number = {3},
    journal = IEEETAES,
    author = {D. C. Woffinden and D. K. Geller},
    year = {2009},
}

@article{GLL18,
    title = {Angles-Only Initial Relative Orbit Determination Algorithm for Non-Cooperative Spacecraft Proximity Operations},
    volume = {2},
    issn = {2522-008X, 2522-0098},
    doi = {10.1007/s42064-018-0022-0},
    pages = {217--231},
    number = {3},
    journal = ASTRODYN,
    author = {B. Gong and W. Li and S. Li and M. Ma and L. Zheng},
    year = {2018},
}

@article{GP15,
    title = {Initial Relative Orbit Determination for Close-in Proximity Operations},
    volume = {38},
    doi = {10.2514/1.G000933},
    pages = {1833--1842},
    number = {9},
    journal = JGCD,
    author = {D. K. Geller and A. Perez},
    year = {2015},
}

@article{DQZ24,
    title={Relative Orbit Determination Algorithm of Space Targets with Passive Observation},
    author={C. Dai and H. Qiang and D. Zhang and S. Hu and B. Gong},
    journal=JSEE,
    volume={35},
    number={3},
    pages={793--804},
    year={2024},
    doi={10.23919/jsee.2024.000051},
}

@article{CX11,
    title={Approach Guidance with Double-Line-of-Sight Measuring Navigation Constraint for Autonomous Rendezvous},
    author={T. Chen and S. Xu},
    journal=JGCD,
    volume={34},
    number={3},
    pages={678--687},
    year={2011},
    doi={10.2514/1.52963},
}

@phdthesis{Cha01,
    title={Autonomous Orbital Rendezvous Using Angles-Only Navigation},
    author={R. J. V. Chari},
    year={2001},
    school={Massachusetts Institute of Technology},
}

@inproceedings{KG12,
    title={Zero $\Delta$v Solution to the Angles-Only Range Observability Problem During Orbital Proximity Operations},
    author={I. Klein and D. K. Geller},
    booktitle=BITZSYM,
    pages={351--369},
    year={2012},
    organization={Springer},
    doi={10.1007/978-3-662-44785-7_19}
}

@article{DBI21,
  author = {F. D'Onofrio and G. Bucchioni and M. Innocenti},
  year = {2021},
  pages = {1-13},
  title = {Bearings-Only Guidance in Cis-Lunar Rendezvous},
  volume = {44},
  journal = JGCD,
  doi = {10.2514/1.G005978}
}

@article{HB83,
  title = {Observability, Eigenvalues, and {Kalman} Filtering},
  author = {F. M. Ham and R. G. Brown},
  year = {1983},
  journal = IEEETAES,
  volume = {AES-19},
  number = {2},
  pages = {269--273},
  issn = {1557-9603},
  doi = {10.1109/TAES.1983.309446}
}

@article{GGL16,
	title={Initial Relative Orbit Determination Analytical Covariance and Performance Analysis for Proximity Operations},
	author={B. Gong and D. K. Geller and J. Luo},
	journal=JSR,
	volume={53},
	number={5},
	pages={822--835},
	year={2016},
	publisher={American Institute of Aeronautics and Astronautics},
	doi={10.2514/1.a33444},
}

@book{BLK01,
	title={Estimation with Applications to Tracking and Navigation: Theory Algorithms and Software},
	author={Y. Bar-Shalom and X. R. Li and T. Kirubarajan},
	year={2001},
	publisher={John Wiley \& Sons}
}

@article{Abd07,
	title={Singular Value Decomposition ({SVD}) and Generalized Singular Value Decomposition},
	author={H. Abdi},
	journal=ENCMEASSTAT,
	volume={907},
	number={912},
	pages={44},
	year={2007},
	publisher={Thousand Oaks (CA) Sage}
}

@article{GL17,
	title={Angles-Only Initial Relative Orbit Determination Performance Analysis Using Cylindrical Coordinates},
	author={D. K. Geller and T. A. Lovell},
	journal=JAS,
	volume={64},
	number={1},
	pages={72--96},
	year={2017},
	publisher={Springer},
	doi={10.1007/s40295-016-0095-z},
}

@article{PGL18,
	title={Non-Iterative Angles-Only Initial Relative Orbit Determination with {$J_2$} Perturbations},
	author={A. C. Perez and D. K. Geller and T. A. Lovell},
	journal=ACTAASTRO,
	volume={151},
	pages={146--159},
	year={2018},
	publisher={Elsevier},
	doi={10.1016/j.actaastro.2018.06.033},
}

@article{EH21,
	title={Review on Relative Navigation Methods of Space Vehicles},
	author={T. Y. Erkec and C. Hajiyev},
	journal=CURRCHINSCI,
	volume={1},
	number={2},
	pages={184--195},
	year={2021},
	publisher={Bentham Science Publishers direct},
	doi={10.2174/2666001601999201210205418},
}

@inbook{BGR24,
  author = {A. S. Booth and R. Gramacy and A. Renganathan},
  title = {Actively Learning Deep {Gaussian} Process Models for Failure Contour and Probability Estimation},
  booktitle = SCITECH,
  doi = {10.2514/6.2024-0577},
  year = {2024},
  publisher = {AIAA},
}

@incollection{SBBHMRTT13,
  author ={M. Saponara and V. Barrena and A. Bemporad and E. N. Hartley and J. Maciejowski 
           and A. Richards and A. Tramutola and P. Trodden},
  title ={Model Predictive Control application to spacecraft rendezvous in {Mars} {Sample \& Return} scenario},
  booktitle=PFDGA,
  volume=6,
  year = 2013,
  pages = {137--158},
  doi = {10.1051/eucass/201306137}
}

@article{DPK12,
  title={Model Predictive Control Approach for Guidance of Spacecraft Rendezvous and Proximity Maneuvering},
  author={S. Di Cairano and H. Park and I. Kolmanovsky},
  journal=IJRNC,
  volume={22},
  number={12},
  pages={1398--1427},
  year={2012},
  doi={10.1002/rnc.2827},
}

@inproceedings{RH03,
  title={Performance Evaluation of Rendezvous Using Model Predictive Control},
  author={A. Richards and J. How},
  booktitle=AIAAGNC,
  pages={5507},
  year={2003},
  doi={10.2514/6.2003-5507},
}

@article{SCG13,
  author={S. Segal and A. Carmi and P. Gurfil},
  title={{Stereovision}-Based Estimation of Relative Dynamics Between Noncooperative Satellites: Theory and Experiments},
  journal=IEEETCST,
  volume={22},
  number={2},
  pages={568--584},
  year={2014},
  doi={10.1109/TCST.2013.2255288},
}

\end{document}